\documentclass{article}

\usepackage[english]{babel}
\usepackage[a4paper,top=2cm,bottom=2cm,left=3cm,right=3cm,marginparwidth=1.75cm]{geometry}
\usepackage{amsmath}
\usepackage{amssymb}
\usepackage{xcolor}
\usepackage{orcidlink}
\usepackage{authblk} 

\newcommand{\bmax}{b_{\text{\scriptsize max}}}
\newcommand{\bcut}{b_{\text{\scriptsize cut}}}
\newcommand{\kcut}{k_{\text{\scriptsize cut}}}
\newcommand{\bmin}{b_{\text{\scriptsize min}}}

\newcommand{\qTeff}{q_T^\text{\scriptsize eff}}
\newcommand{\ordof}[1]{\mathcal{O}\big(#1\big)}

\newcommand{\lambdaQCD}{\Lambda_{\text{\tiny QCD}}}
\newcommand{\fourier}[1]{\widetilde{#1}}
\newcommand{\mellin}[1]{\hat{#1}}
\newenvironment{acknowledgements}{
  \section*{Acknowledgements}
}{}

\title{
Transverse momentum Resummation \\and Analytic continuation into the Deep Infrared
}

\author[1]{Andrea Simonelli\orcidlink{0000-0003-2607-9004}\thanks{e-mail: andrea.simonelli@roma1.infn.it}}
\affil[1]{INFN, Sezione di Roma, Piazzale Aldo Moro 5, 00185 Roma, Italy}

\date{September 2025}

\begin{document}
\maketitle

\begin{abstract}
Extracting information on the three-dimensional structure of hadrons from transverse-momentum data is a challenging task, requiring the separation of perturbative contributions under QCD control from genuinely non-perturbative effects, unavoidably affected by phenomenological bias. 
Standard approaches model transverse-momentum–dependent (TMD) observables in impact-parameter space and compare them with data after Fourier inversion, generating an interplay between perturbative terms, non-perturbative models, and prescriptions used to extend perturbation theory beyond its natural domain.
In this work, a consistent methodology to perform the Fourier inversion analytically is proposed, achieving a solid resummation of TMD parton distributions and cross sections directly in transverse-momentum space. 
The resummation is carried out to next-to-leading logarithmic (NLL) accuracy, validated against alternative prescriptions, and tested on experimental data. This framework provides a systematic procedure to analytically continue observables into the deep infrared,  identifying two primary sources of non-perturbative effects: the low energy behavior of the strong coupling and of the parton distribution functions.
This points to a paradigm shift in TMD phenomenology, where the structure of hadrons emerges from the dynamics of QCD itself rather than from flexible parametrizations with only indirect physical interpretation.
\end{abstract}

\section{Introduction}
\label{sec:intro}

Transverse momentum dependent observables provide sharp probes of the internal structure of hadrons, as they are directly sensitive to the three-dimensional motion of confined partons. Measuring the transverse momentum $q_T$ of a final state with respect to a well-defined axis enlarges the range of accessible kinematic configurations. Depending on the relative size between $q_T$ and the hard scale $Q$, the factorization properties of the process change significantly. 
For $q_T \approx Q$, the transverse-momentum dependence is fully perturbative and contained into a partonic coefficient function, while all nonperturbative effects are encoded in parton distribution functions (PDFs) and/or fragmentation functions (FFs) depending on the process. 
A relevant example is the (unpolarized) Drell-Yan cross section: 
\begin{align}
    \label{eq:collinear_fact}
    &\frac{d \sigma}{d Q^2 \, d y \,d q_T^2}
    =
    \sum_{i,j} \tau \int_{\tau}^1 \frac{d x}{x} \int d \eta \frac{d C_{i j}\big(a_S(\mu),x,\eta;{Q}/{\mu}\big)}{d Q^2 d \eta d q_T^2} 
    \notag \\
    &\qquad\times
    f_{i}\big(\sqrt{\tau/x} e^{y-\eta},\mu\big)
    f_{j}\big(\sqrt{\tau/x} e^{-(y-\eta)},\mu\big)
    \theta\big(e^{-2 |y-\eta|} - \tau/x\big)
\end{align}
where $Q^2$, $y$, and $\vec{q}_T$ are the invariant mass, the rapidity in the center-of-mass, and the transverse momentum with respect to the beam direction of the Drell-Yan pair. The sum over the two colliding hadrons, as well as polarization effects, is implicit and not shown for clarity. The ratio $\tau = Q^2/S$ represents the relative size of $Q$ compared to the center-of-mass energy $\sqrt{S}$, while $x$ and $\eta$ are the partonic analogues of $\tau$ and $y$, respectively. The sum runs over all partonic species, quark, antiquarks and gluons, initiating the hard scattering.

In contrast, when $q_T \ll Q$, the transverse-momentum dependence cannot be described by a fixed-order expansion in the strong coupling and non-perturbative effects are encoded in TMD parton densities, three-dimensional generalizations of PDFs and FFs:
\begin{align}
    \label{eq:tmd_fact}
    &\frac{d \sigma}{d Q^2 \, d y \,d q_T^2}
    =
    H\big(a_S(\mu);{Q}/{\mu}\big) 
    \int d^2\vec{k}_T d^2\vec{k}'_T \,
    \delta\big(\vec{q}_T - \vec{k}_T - \vec{k}'_T\big)
    \notag \\
    &\qquad\times
    \sum_q e_q^2
    F_{q}\big(\sqrt{\tau} e^{y}, k_T; \mu, \zeta/\mu^2\big)
    F_{\overline{q}}\big(\sqrt{\tau} e^{-y}, k_T'; \mu, \zeta/\mu^2\big)
\end{align}
where $\zeta \approx Q^2 e^{-2 y_n}$ is the rapidity scale induced by the off-lightcone tilt\cite{Collins:2011zzd,Simonelli:2025xpm} $y_n$ that characterize TMD operators. This time, the sum runs over quarks and antiquarks, as gluon-initiated partonic scattering are suppressed in the low transverse momentum region. As before, hadron sums and polarization effects are implicit.

The shift from Eq.\eqref{eq:collinear_fact} to Eq.\eqref{eq:tmd_fact} originates from the gradual replacement of hard gluon emissions by soft and collinear radiation as $q_T$ decreases, thereby modifying the leading momentum regions of the process. As a result, large logarithms of the form $\log{(Q/q_T)}$ appear at all orders in perturbation theory, and spoil the convergence of the fixed-order expansion. 
The treatment of this regime is delicate, and its difficulties have been recognized since the early days of QCD. It soon became clear that mapping the observables to the Fourier-conjugate space and expressing them in terms of the impact parameter $b_T$ is particularly convenient: not only does it make the analytic structure transparent\cite{Parisi:1979se}, but it also renders factorization manifestly simple\cite{Collins:1981uk,Collins:1981va,Collins:1984kg,Collins:2011zzd}, turning convolutions in transverse momentum space into ordinary products. In this auxiliary space, momentum conservation and the vector nature of $\vec{q}_T$ are fully accounted for by the Fourier factor, so they do not obstruct the resummation of large logarithms, which yields the well-known Sudakov factor. 

Nevertheless, impact-parameter space has its limitations\cite{Ellis:1997ii,Ellis:1997sc}. Chief among them is that it constitutes an auxiliary space, distinct from the transverse momentum space in which measurements are performed, and this detour may obscure the physical interpretation of the results. In $b_T$-space, the disentangling of non-perturbative effects is inevitably blurred by the inverse Fourier transform, which poses a risk when comparing with experimental data. In addition, the matching of the two factorization theorems at intermediate $q_T$ relies on delicate cancellations between their asymptotics, potentially distorted by the Fourier inversion. 

These limitations are generally regarded as acceptable compromises in exchange for analytic control and clean factorization. Formulations in $q_T$-space, by contrast, are notoriously difficult. Purely analytic results suffer from spurious singularities\cite{Frixione:1998dw}, while series expansions\cite{Kulesza:1999gm,Kulesza:1999sg,Kulesza:2001jc} depend sensitively on the truncation point and tend to exacerbate the issues of the matching region at large $q_T$. Over the years, increasingly refined resummation techniques have been proposed\cite{Ebert:2016gcn}, together with semi-analytical results\cite{Kang:2017cjk}. Yet, a fully analytic closed-form expression remains unavailable beyond leading-logarithmic (LL) accuracy, whose classic formulation goes back to Ellis and Veseli\cite{Ellis:1997sc}. 
Indeed, when working directly in $q_T$-space, most approaches abandon analyticity altogether and rely on coherent parton branching techniques\cite{Banfi:2004yd,Banfi:2014sua,Monni:2016ktx}.

In this work, we propose a new analytic formulation for $q_T$ resummation that builds on earlier approaches\cite{Ellis:1997sc,Kulesza:1999gm,Kulesza:1999sg,Kulesza:2001jc,Frixione:1998dw} and consistently extends the result of Frixione, Nason, and Ridolfi\cite{Frixione:1998dw}, achieving next-to-leading-logarithmic (NLL) accuracy for both TMDs and TMD-factorized cross sections.
The analytic formulation proposed here aims at extracting the maximum insight from perturbative QCD, introducing a novel strategy to access the deep infrared through a controlled analytic continuation of perturbative results. This approach makes it possible to transparently identify the sources of non-perturbative effects, tracing them back directly to the infrared behavior of the strong coupling and of the parton distributions.
Efforts in the same spirit, emphasizing the physical interpretation of the distributions, have also been pursued in the works of Rogers and collaborators\cite{Gonzalez-Hernandez:2022ifv,Gonzalez-Hernandez:2023iso}, albeit following a rather different, and somehow opposite, path: there, the perturbative content is built on top of infrared models, whereas here the logic is reversed.
Both perspectives stand in contrast with the standard treatments in TMD physics, where non-perturbative contributions are typically absorbed into highly flexible auxiliary functions, with a looser connection to the underlying QCD dynamics.

To date, a consistent and comprehensive extraction of collinear PDFs within the TMD factorization framework is still lacking, even though $q_T$-dependent data are inherently more sensitive to parton dynamics than inclusive measurements.
The formulation developed in this work represents a step in this direction, since collinear PDFs become actual cornerstones of TMD phenomenology, which emerges as a sharp probe of the deep infrared regime of QCD. Beyond the study of non-perturbative physics, this approach has the potential to significantly reduce theoretical uncertainties on PDFs, allowing them to be probed in extreme kinematic regions, such as very small or very large $x$, and at low energy scales, where they remain most elusive.

\section{Impact parameter space}
\label{sec:bT-space}

In impact parameter space, the problematic region of low transverse momentum is mapped to large distances, $b_T \gg 1/Q$. In this regime, large logarithms, this time of the form $\log(b_T Q/c_1)$ with $c_1 = 2 e^{-\gamma_E} \approx 1.12\dots$, emerge and undermine the convergence of the perturbative expansion. This emerges already at the first non-trivial order of TMD distributions:
\begin{align}
    \label{eq:tmd_NLO}
    &F_j\big(x,b_T,\mu,\zeta/\mu^2\big)
    =
    f_j(x;\mu) 
    + a_S(\mu) \,
    \int_x^1 \frac{d \hat{x}}{\hat{x}}
    \sum_k \Big[
    C^{[1]}_{j/k}(\hat{x}) 
    -2 L_b P_{j/k}^{[0]}(\hat{x}) 
    \notag \\
    &\;+
    \big[
    L_b \gamma_f^{[0]} - \frac{1}{2} \gamma_K^{[0]} (L_b^2 + L_b L_\zeta)
    \big]
    \delta(1-\hat{x}) \delta_{j k}
    \Big]
    \,f_k(x/\hat{x};\mu) + \ordof{a_S^2(\mu)}
\end{align}
where, at the TMD level, $\mu$ replaces $Q$, and we have defined $L_b = \log(b_T \mu/c_1)$ and $L_\zeta = \log(\zeta/\mu^2)$. Here, $P_{j/k}$ denote the usual DGLAP splitting kernels\cite{Moch:2004pa,Vogt:2004mw}, $\gamma_f$ and $\gamma_K$ are the TMD\cite{Collins:2017oxh} and cusp anomalous dimensions\cite{Henn:2019swt}, respectively, and $C_{j/k}$ represent\footnote{Note that the notation $C_{j/k}$ commonly refers to the full function convoluted with the PDF, not just the non-logarithmic component.} perturbatively calculable coefficient functions\cite{Collins:2017oxh}, determined order by order in ${a_S = \alpha_S/(4\pi)}$. 
Relevant values are collected in Appendix \ref{app:resummation}.
The sum starting at NLO runs over all parton species. This expression is an operator product expansion (OPE) onto collinear distributions and, as long as $b_T \approx c_1/\mu$, it provides a reliable estimate of the TMD operator. 

This description, however, gradually breaks down at larger $b_T$, as the logarithms $L_b$ grow to a size that compensates for the smallness of $a_S(\mu)$.  The logarithms appearing in Eq.\eqref{eq:tmd_NLO} can be formally resummed to all orders quite simply, owing to the structure of TMD evolution equations in $b_T$ space:
\begin{subequations}
\label{eq:TMD_evo}
    \begin{align}
     &\frac{\log{F_j\big(x,b_T,\mu,\zeta/\mu^2\big)}}{d \log{\mu}}
    =
    \gamma_F\big(a_S(\mu), {\zeta}/{\mu^2}\big);
    \label{eq:TMD_evo_mu}
    \\
    &\frac{\log{F_j\big(x,b_T,\mu,\zeta/\mu^2\big)}}{d \log{\sqrt{\zeta}}}
    =
    K\big(a_S(\mu),{b_T \mu}/{c_1} \big),
    \label{eq:TMD_evo_K}
    \end{align}
\end{subequations}
together with the renormalization group flow of the Collins-Soper (CS) kernel $K$:
\begin{align}
    \label{eq:K_evo}
    \frac{d K\big(a_S(\mu), {b_T \mu}/{c_1}\big)}{d \log{\mu}} = \gamma_K\big(a_S(\mu)\big).
\end{align}
The solution to the equations above is readily obtained:
\begin{align}
    &F_j(x,b_T;\mu,\zeta/\mu^2) = F_j(x,b_T;\mu_b,1) 
    \notag \\
    &\times
    \text{exp}\Big\{
    L_b K\big(a_S(\mu_b); 1\big) 
    +
    \int_{\mu_b}^\mu \frac{d \mu'}{\mu'}
    \Big[
    \gamma_f\big(a_S(\mu')\big)
    -\gamma_K\big(a_S(\mu')\big)\,
    \log{\big({\mu}/{\mu'}\big)}
    \Big]
    \Big\}
    \notag \\
    &\times
    \text{exp}\Big\{
    \frac{1}{2} L_\zeta K\big(a_S(\mu); b_T \mu/c_1\big) 
    \Big\}
    \label{eq:tmd_solEVO}
\end{align}
where the reference scales have been appropriately set to $\mu_b = c_1/b_T$ and $\zeta_b = \mu_b^2$. These are the scales that optimize the convergence of the perturbative expansion, by moving all the dangerous logarithms into the exponent. 

\begin{figure}
    \centering
    \includegraphics[width=0.75\linewidth]{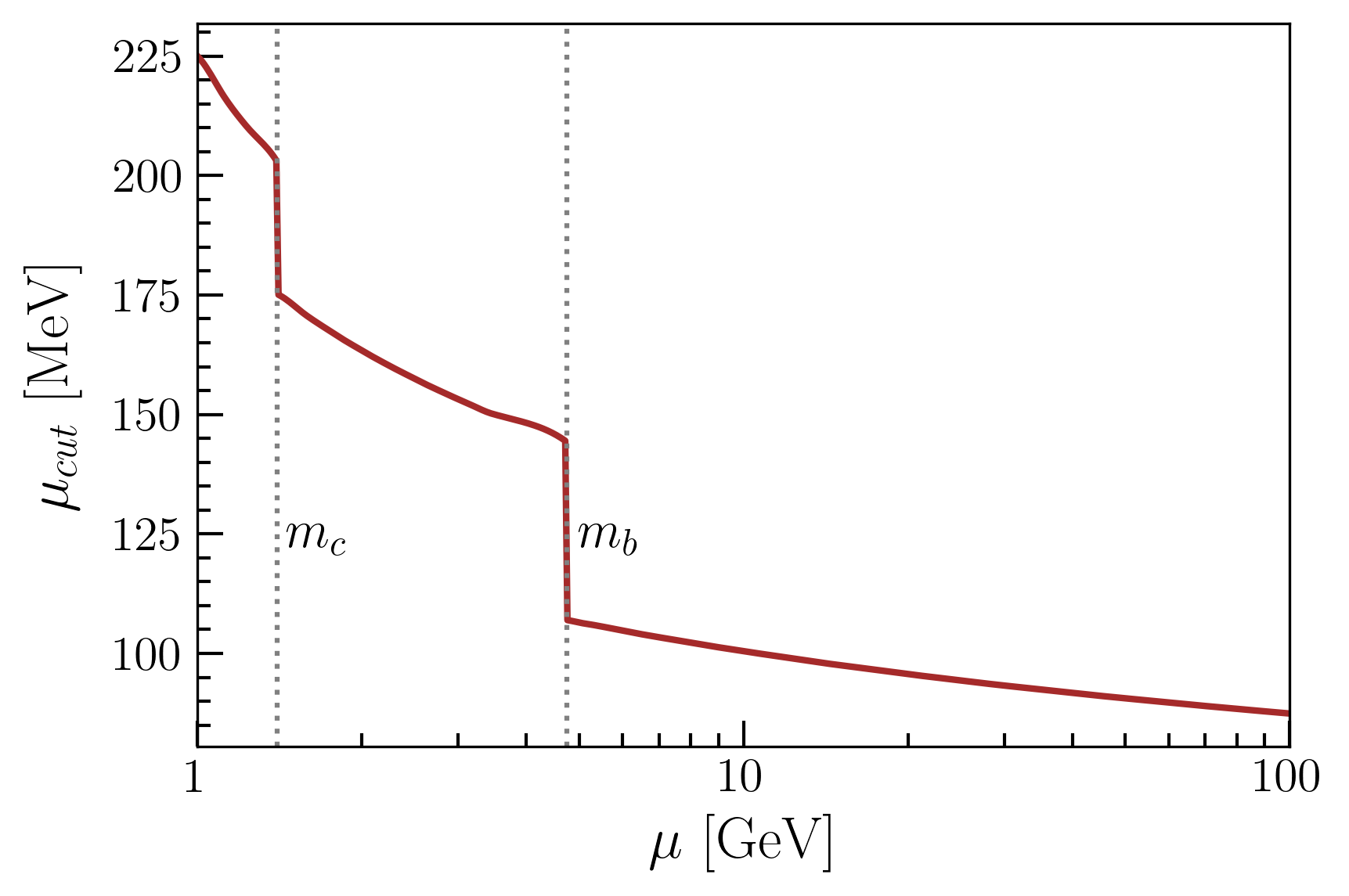}
    \caption{Energy scale behavior associated with the Landau pole identified by $\mu_{\text{\footnotesize cut}} = c_1/\bcut(\mu)$. Vertical dotted lines indicate heavy-quark mass thresholds, where the QCD beta function exhibits discontinuities. The value of alpha strong and the quark masses are taken from MSHT20 PDF set at NNLO\cite{Bailey:2020ooq}.}
    \label{fig:bcut}
\end{figure}

Provide the scale $\mu_b$ is large enough for perturbation theory to apply, roughly above 1-2 GeV, the integral over the anomalous dimension can be carried out analytically, yielding functions of $a_S(\mu)$ and $a_S(\mu_b)$:
\begin{align}
    \label{eq:anom_dim_int}
    &\int_{\mu_b}^\mu \frac{d \mu'}{\mu'}
    \Big[
    \gamma_f\big(a_S(\mu')\big)
    -\gamma_K\big(a_S(\mu')\big)\,
    \log{\big({\mu}/{\mu'}\big)}
    \Big]
    \notag \\
    &\quad=
    -\frac{\gamma_K^{[0]}}{4\beta_0^2}\Big[
    \frac{1}{a_S(\mu_b)} - \frac{1}{a_S(\mu)}\Big(1 + \log{\frac{a_S(\mu)}{a_S(\mu_b)}}\Big)
    \Big] + \dots
\end{align}
and the functions at the reference scale can be safely expanded in powers of $a_S(\mu_b)$:
\begin{subequations}
\label{eq:tmd_RefScale}
    \begin{align}
    &F_j(x,b_T;\mu_b,1) =
    f_j(x;\mu_b) + a_S(\mu_b) \, 
    \sum_k \int_x^1 \frac{d\hat{x}}{\hat{x}}
    C^{[1]}_{j/k}(\hat{x}) f_k(x/\hat{x};\mu_b) + \dots;
    \\
    &K\big(a_S(\mu_b); 1\big)  =
    a_S^2(\mu_b) \, k^{[2]} + \dots,
    \end{align}
\end{subequations}
where the coefficients $k$ are known at NNLO\cite{Collins:2017oxh}.
Extending these expression below $\mu_b \lesssim 1$–$2$ GeV would require knowledge of both the strong coupling and the PDFs in the deep infrared regime of QCD. In fact, at such low scales, these approximations may cease to be valid altogether and be replaced by a different functional behavior, beyond the reach of perturbation theory. 
Within the perturbative domain, however, the expansions can be conveniently recast in terms of $a_S(\mu)$ and powers of $L_b$, by evolving $a_S(\mu_b)$ with the renormalization group:
\begin{align}
    \label{eq:aSmub_RG}
    a_S(\mu_b) = \frac{a_S(\mu)}{1-\lambda_b}
    \, \Big[
    1 - a_S(\mu) \frac{\beta_1}{\beta_0}
    \frac{\log{(1 - \lambda_b)}}{1-\lambda_b}
    +
    \dots
    \Big]
\end{align}
where we have set $\lambda_b = 2 \beta_0 a_S(\mu) L_b$ and $\beta_i$ are the coefficients of the QCD beta function. This leads to:
\begin{align}
    &F_j(x,b_T;\mu,\zeta/\mu^2) = 
    \Big[f_j(x;\mu_b) + \dots\Big]\,
    \text{exp}\Big\{
    L_b g_1(\lambda_b) + g_2(\lambda_b) + \dots
    \Big\}
    \notag \\
    &\times
    \text{exp}\Big\{
    \frac{1}{2} L_\zeta \big[
    \kappa_1(\lambda_b) + \frac{1}{L_b} \kappa_2(\lambda_b) + \dots \big]
    \Big\},
    \label{eq:tmd_bT_resummed}
\end{align}
where the $\dots$ stand for contributions that are at least of order $\mathcal{O}(1/L_b)$ or $\mathcal{O}(L_\zeta/L_b^2)$. The functions $g_i$ and $\kappa_i$ (see Appendix \ref{app:resummation}) depend on the anomalous dimensions and become non-analytic at $\lambda_b = 1$, where a branch cut develops.
Away from this point, the expression can be expanded in powers of $a_S(\mu)$, reproducing Eq.\eqref{eq:tmd_NLO}.
In this expansion, each inverse power of $L_b$ resums an infinite tower of logarithms: the leading tower corresponds to the leading-logarithmic (LL) series $a_S^n(\mu) L_b^{2n}$ generated by $g_1$, the next-to-leading tower to the next-to-leading logarithmic (NLL) series $a_S^n(\mu) L_b^{2n-1}$ generated by $g_2$, and so forth.
As a result, the validity of the expression extends well beyond the region $b_T \approx c_1/\mu$, naively up to $\lambda_b \lesssim 1$. 
The critical value at the branch point $\bcut(\mu) = c_1/\mu \, \text{exp}\big\{{1}/{(2\beta_0 a_S(\mu))}\big\}$ is the manifestation of the Landau pole at large distances. 
This scale is extremely low, of the order of $100$–$200$ MeV, even below $\Lambda_\text{QCD}$, as illustrated in Fig.~\ref{fig:bcut}. 
Therefore, it cannot be interpreted as a practical boundary between perturbative and non-perturbative regimes.
Nevertheless, as this threshold is approached, Eq.~\eqref{eq:tmd_bT_resummed} becomes increasingly unreliable as an estimate of the TMD operator.
A more realistic transition to the non-perturbative regime is expected at a scale $Q_0 \sim 1$–$2$ GeV, roughly corresponding to the proton mass or the chiral-symmetry-breaking scale. Below this scale, non-perturbative corrections must be explicitly introduced.

In impact-parameter space, this step is unavoidable before transforming back to transverse momentum space, since the inverse Fourier transform runs over all values of $b_T$, from short to very large distances. As a result, the TMD distributions must be specified across the entire range of $b_T$, not just below $\bcut$ (or, more realistically, below $c_1/Q_0$). 
This is a crucial feature of the auxiliary space, as any comparison with experimental data is inevitably filtered through the inverse Fourier transform, which mixes contributions from short and large distances. Defining and implementing models in $b_T$-space is therefore particularly delicate, as one must carefully preserve the physical meaning of contributions from different distance regimes.
In $b_T$-space, the boundaries are reasonably well-defined: around $b_T \approx c_1/\mu$ , the fixed-order expression of Eq.\eqref{eq:tmd_NLO} is valid; in the intermediate region
$ c_1/\mu \lesssim b_T \lesssim c_1/Q_0$, the resummed expression of Eq.\eqref{eq:tmd_bT_resummed} provides a reliable estimate; and for $b_T \gtrsim c_1/Q_0$ non-perturbative corrections must be included.
However, once the distributions are transformed back to $q_T$-space, these boundaries become blurred: non-perturbative contributions can spill into the intermediate region, and the intermediate region itself, where perturbative methods remain valid but large logarithms dominate, may not be properly captured.

\bigskip

Since the Landau pole occurs at distances much larger than the scales where non-perturbative effects are expected to become relevant, the issue of avoiding it through a perturbative prescription is effectively absorbed into the broader problem of modeling the infrared behavior of TMD distributions. 
Both the Landau-pole–avoiding prescription and the introduction of non-perturbative models, when applied to Eq.~\eqref{eq:tmd_bT_resummed}, aim to extend calculations towards low-energy scales, yet they are fundamentally different in nature.
Prescriptions to avoid the Landau pole\cite{Collins:1981va,Catani:1996yz,Kulesza:2002rh,Bonvini:2008ei} are artificial devices, designed to formally prolong perturbative calculations beyond their natural domain. By contrast, non-perturbative models provide genuine physical input, enabling a consistent description of operators and cross sections across all scales.

The typical strategy, rooted in the seminal works of Collins, Soper, and Sterman \cite{Collins:1984kg} and adopted in most modern TMD extractions \cite{Bacchetta:2022awv,Moos:2023yfa,Bacchetta:2024qre,Moos:2025sal,Bacchetta:2025ara,Camarda:2025lbt}, introduces a Landau-pole–avoiding prescription by replacing the $b_T$-dependence in Eq.\eqref{eq:tmd_solEVO} with a functional form $b_T^\star$. This function coincides with $b_T$ at short distances and saturates at a scale $\bmax \lesssim \bcut$ at large distances, thereby ensuring that the Landau singularity is never reached. The scale $\bmax$ effectively sets the boundary of perturbative validity: both the strong coupling $\alpha_S$ and the PDFs are evaluated at scales not lower than $Q_0 = c_1/\bmax$. Note that this procedure effectively introduces a saturating-like non-perturbative modeling of both the strong coupling and the parton distributions, in contrast with the original idea of employing a mere technical prescription to extend perturbative calculations. In fact, in the original formulation the $\bmax$-dependence is supposed to cancel once suitable multiplicative non-perturbative factors are introduced in the TMD distribution. However, this cancellation is only formal: in practice, it is virtually impossible to construct non-perturbative corrections with the required $\bmax$-dependence to achieve an exact cancellation. 
The scale $\bmax$ thus becomes a genuine parameter, intrinsically and unavoidably correlated with the physical inputs introduced in non-perturbative models that shape the infrared behavior of both TMDs and cross sections. Its impact is particularly severe in the intermediate region, an effect only partially mitigated by including data across multiple energy scales (global fits)\cite{Cerutti:2025inprep}.

This is not necessarily a drawback, as long as one acknowledges that the $b^\star$ prescription is more than a mere technical device and effectively forms part of the non-perturbative model. Nevertheless, this raises the question of whether the commonly adopted scale $Q_0 \approx 1$–$2$ GeV is really adequate to describe the saturation of $\alpha_S$ in the deep infrared regime of QCD. It seems more natural to expect a scale of the order of $\lambdaQCD$. This implies that the saturation scale historically tied to $\bmax$ should be carefully disentangled from the perturbative/non-perturbative interface scale $Q_0$, which has often been conflated in practical extractions. 
Similar arguments equally hold for the collinear PDFs contributing to the TMDs, as thoughtfully discussed in Section\ref{sec:deepIR}. 
Such a disentanglement has also been advocated in Refs.\cite{Gonzalez-Hernandez:2022ifv,Gonzalez-Hernandez:2023iso}, where $Q_0$ is indeed treated as the input scale at which the non-perturbative content is defined.
Such efforts would encourage the development of models reflecting the fundamental features of QCD, rather than highly flexible parametrizations devised to absorb our ignorance of the infrared behavior of TMD operators.

This ambitious plan is difficult to pursue in $b_T$-space, due to the natural reshuffling of contributions induced by the inverse Fourier transform. It would be far more practical to carry out the program entirely in $q_T$-space, where the different scales and models can be tested directly against experimental data. Moreover, the matching between the two regimes of Eq.\eqref{eq:collinear_fact} and Eq.\eqref{eq:tmd_fact} would be more natural if both expressions were evaluated in transverse momentum space, avoiding the artificial oscillations induced by the inverse Fourier transform at large $q_T$. However, achieving in $q_T$-space the same level of analytic insight as in $b_T$-space is more challenging, and it will be the focus of the next Section.

\section{Transverse Momentum Space}

A direct physical intuition and a transparent implementation of matching provide strong motivations to pursue a resummation formalism formulated directly in $q_T$-space, through an analytic inversion of the Fourier transform, while preserving the logarithmic accuracy that characterizes the $b_T$-space approach.
Unfortunately, this program encounters fundamental obstacles already at next-to-leading logarithmic (NLL) accuracy\cite{Frixione:1998dw}. The difficulty stems from the vector nature of transverse momentum: configurations with large individual partonic transverse momenta can combine, through azimuthal cancellations, into a small net $q_T$ (or $k_T$ at the level of TMD distributions). These highly energetic emissions eventually disrupt the Sudakov resummation in $q_T$-space: the dangerous large logarithms $\log(Q/q_T)$ do not consistently exponentiate so as to reproduce the exact analytic counterpart of Eq.\eqref{eq:tmd_bT_resummed}. 
But their impact is far more severe: left untreated, they give rise to spurious poles at low $q_T$, non-physical artifacts of the formalism that fatally compromise any predictive intent. 
Two main strategies have been proposed: one circumvents the problem at its root, by carrying out the resummation with respect to the hardest partonic emission instead of $q_T$, as in \verb|RADISH| framework\cite{Monni:2016ktx,Bizon:2017rah,Re:2021con,Buonocore:2024xmy}; the other addresses it directly by solving the TMD evolution equations of Eqs.\eqref{eq:TMD_evo} in $q_T$-space in terms of distributions\cite{Ebert:2016gcn}. Both formulations result in infinite convolutions. While they correctly account for the highly energetic contributions and avoid unphysical pathological spurious poles, they do not provide closed analytic expressions. This can potentially obscure the physical intuition behind the non-perturbative corrections that we aim to capture.

The dangerous contributions originate from the low-$b_T$ region, which corresponds to large partonic transverse momenta. Since the inverse Fourier transform integrates over the entire $b_T$ spectrum, it inevitably includes this problematic domain. Frixione, Nason, and Ridolfi proposed\cite{Frixione:1998dw} to exclude this region by introducing a scale $\bmin = c_1/Q$, below which the cross section is not evaluated. 
In the following, we demonstrate how to confront these dangerous contributions directly, obtaining a closed analytic expression without the need for any additional scale.

\subsection{Analytic Fourier Inversion}
\label{sec:analytic_FT}

The analytic Fourier inversion technique presented here maps a function defined in impact parameter space, approximated for $b_T \lesssim c_1/Q_0$ as in Eq.\eqref{eq:tmd_bT_resummed}, into its transverse momentum-space counterpart, valid for $k_T \gtrsim Q_0$, while preserving the same logarithmic accuracy. Note that the Sudakov-style logarithmic counting is only strictly maintained at leading logarithmic (LL) accuracy; beyond that, the main feature retained is that the resulting expression reproduces the intended log counting when expanded in powers of alpha strong. The transverse momentum–space representation of the OPE in Eq.~\eqref{eq:tmd_NLO} reads
\begin{align}
\label{eq:tmd_NLO_kTspace}
    &F_j(x,k_T,\mu,\zeta/\mu^2) = 
    \frac{a_S(\mu)}{2\pi k_T^2}
    \int_x^1 \frac{d \hat{x}}{\hat{x}}
    \sum_k 
    \Big[
    2 P^{[0]}_{j/k}(\hat{x})-
    \notag \\
    &\quad
    - \big[ 
    \gamma_f^{[0]} 
    - \frac{1}{2}\gamma_K^{[0]}\log{\big({\zeta}/{k_T^2}\big)}
    \big] \delta(1-\hat{x})
    \delta_{j k}
    \Big]
    f_k(x/\hat{x};\mu)
    +
    \ordof{a_S^2(\mu)}\,.
\end{align}
At NLL accuracy, the resummed expression must reduce exactly to this fixed-order result. 
The strategy is first demonstrated on the TMD distribution, where $k_T$ and $\mu$ play the roles of $q_T$ and $Q$, before naturally extending the same approach to the TMD-factorized cross section of Eq.\eqref{eq:tmd_fact}.
Throughout this Section, we suppress all arguments except for $x$ and $k_T$ (or $b_T$) for notational simplicity, and quantities in impact parameter space are denoted with a tilde to distinguish them from their transverse-momentum counterparts. Also, we set $\zeta = \mu^2$ at level of TMDs, since this choice does not affect the derivation and the rapidity-scale contribution (the last exponential in Eq.\eqref{eq:tmd_bT_resummed}) cancels out at the level of the cross section. An alternative choice, such as the so-called 
$\zeta$ prescription \cite{Scimemi:2017etj,Scimemi:2018xaf}, does not alter the general conclusions of this Section or of the present work. The explicit $\zeta$-dependence can be retained in the distributions in transverse momentum space, as discussed in Appendix \ref{app:zeta_rap}.

\bigskip

The TMD distribution $F_j(x,k_T)$ can be conveniently defined in transverse momentum space using a trick commonly employed in other resummation frameworks, such as for event-shape observables\cite{Catani:1992ua}. The starting point is the cumulative distribution
\begin{align}
R_j(x,k_c) = \int d^2 \vec{k}_T \, F_j(x,k_T)\, \theta(k_c - k_T),
\end{align}
which corresponds to the TMD integrated up to a cutoff $k_c$ (or its zero-moment\cite{delRio:2024vvq}). This quantity is clearly related to the collinear counterpart of $F_j$, and it has recently been proposed\cite{Gonzalez-Hernandez:2022ifv} as an alternative renormalization scheme for PDFs. Its relation to the standard $\overline{\text{MS}}$-scheme scheme can be expressed as a perturbatively stable series:
\begin{align}
    \label{eq:cutoff_scheme}
    &R_j(x,\mu) = f_j(x;\mu) + a_S(\mu) \sum_k \big[C^{[1]}_{j/k}\otimes f_k\big](x;\mu) + \ordof{a_S^2(\mu)}
\end{align}
where the convolutions involve the OPE coefficients that are not associated with logarithmic powers and the sum runs over all partonic species.
The TMD is then obtained as a derivative with respect to the cutoff:
\begin{align}
F_j(x,k_T) = \frac{1}{2\pi k_T^2} \left. \frac{d R_j(x,k_c)}{d \log k_c} \right|_{k_c = k_T}.
\end{align}
Substituting the Fourier representation of the TMD,
\begin{align}
\label{eq:tmd_fourier}
F_j(x,k_T) = \int \frac{d^2 \vec{b}_T}{(2\pi)^2} \, e^{i \vec{k}_T \cdot \vec{b}_T} \fourier{F}_j(x,b_T),
\end{align}
into the cumulative distribution, and exploiting rotational symmetry, the two-dimensional Fourier integral reduces to a one-dimensional Bessel transform:
\begin{align}
\label{eq:Rc_recasted}
R_j(x,k_c) = \int_0^\infty d\xi \, J_1(\xi) \, \fourier{F}_j(x, \xi/k_c),
\end{align}
where $J_1$ is the Bessel function of the first kind and the rescaled variable $\xi = k_c b_T$ has been introduced.
The TMD can now be written as in Eq.\eqref{eq:tmd_bT_resummed}, with the substitution $L_b = L_c + L_\xi$, where $L_c = \log(\mu/k_c)$ and $L_\xi = \log(\xi/c_1)$. The Sudakov factor, i.e. the exponent in Eq.\eqref{eq:tmd_bT_resummed} organized as an expansion in inverse powers of $L_b$, can be equivalently recasted as:
\begin{align}
    \label{eq:Sudakov_recasted}
    &S(L_c + L_\xi) =
    \notag \\
    &\quad=
    S(L_c) + L_\xi H_1\big(\lambda_c, {L_\xi}/{L_c}\big)
    + H_2\big(\lambda_c, {L_\xi}/{L_c}\big)
    + \frac{1}{L_\xi} H_3\big(\lambda_c, {L_\xi}/{L_c}\big) + \dots
\end{align}
where each function $H_i$ depends solely on the corresponding N$^i$LL $g_i$ function and is therefore fully determined once the target logarithmic accuracy is specified:
\begin{subequations}
\label{eq:Hi_functions}
\begin{align}
H_1(\lambda,\rho) &= 
\sum_{k\geq0} \frac{1}{(1+k)!} h_1^{(k)}(\lambda) (\lambda \, \rho)^k , &
h_1^{(k)} &= \frac{d^k}{d \lambda^k} \Big[ g_1(\lambda) + \lambda g'_1(\lambda) \Big] \,,
\label{eq:H1} \\
H_2(\lambda,\rho) &= 
\sum_{k\geq1} \frac{1}{k!} h_2^{(k)}(\lambda) (\lambda \, \rho)^k , &
h_2^{(k)} &= \frac{d^k}{d \lambda^k} g_2(\lambda) \,,
\label{eq:H2} \\
H_3(\lambda,\rho) &= 
\rho^2 \sum_{k\geq0} \frac{1}{(1+k)!} h_3^{(k)}(\lambda) (\lambda \, \rho)^k , &
h_3^{(k)} &= L^2 \frac{d^k}{d \lambda^k} \Big[ \frac{-g_3(\lambda) + \lambda g'_3(\lambda)}{L^2} \Big] 
\label{eq:H3}
\end{align}
\end{subequations}
and similarly for higher-order functions $H_4$, $H_5$, etc.
The logarithmic accuracy defined at the level of the integrand must be carefully propagated through the integration, keeping in mind that the logarithmic counting in $L_c$ is entangled with that of $L_\xi$, which is an integration variable: in certain regions of phase space, the largeness of $L_c$ can be compensated by the largeness of $L_\xi$. 
As a general recipe, to preserve the logarithmic accuracy through the integration, it is sufficient to expand the first term $S(L_c)$ to the desired inverse power of $L_c$, while keeping a consistent amount of powers of $L_\xi$ in the expansion in terms of the functions $H_i$: $\ordof{L_\xi^0}$ at LL, $\ordof{L_\xi^1}$ at NLL, and so on. 

The OPE of the TMD distributions at reference scale, i.e., the terms not exponentiated in Eq.\eqref{eq:tmd_bT_resummed}, can be treated similarly, with the important caveat that the collinear PDFs are now evaluated at the scale $\mu = ({c_1}/{\xi}) \, k_c$. Their evolution to the scale $k_c$ can be performed using DGLAP.
Since the logarithmic structure is more transparent in Mellin space, it is convenient to perform the evolution there. This allows for full analytic control of the resummation and for a systematic counting of logarithms. In full generality\cite{Simonelli:2024vyh}, one has\footnote{The flavor dependence is not relevant here, both singlet and non-singlet sector share the same analytic structure.}:
\begin{align}
\label{eq:DGLAP_evo}
&\mellin{f}\big(N;({c_1}/{\xi}) \, k_c\big)
=
\text{exp}\Big\{ \big[
-\frac{1}{\beta_0}\log{(1+\lambda_\xi)} + \dots 
\big] \mellin{P}_0(N) \Big\}
\times \dots \times
\mellin{f}(N,k_c),
\end{align}
where $\lambda_\xi = 2 \beta_0 a_S(\mu) L_\xi$ and $\mellin{f}$, $\mellin{P}_i$ are the Mellin transforms of the PDFs and splitting functions, respectively. 
In this expression, the $\dots$ denote terms suppressed at least as $\ordof{1/L_\xi}$. Such contributions are reabsorbed by the $H_2$ term (and higher) in Eq.~\eqref{eq:Sudakov_recasted}. As a consequence, explicit DGLAP splitting kernels enter the calculation only from N$^2$LL onward. Thus, up to NLL accuracy, the PDFs in the integrand can be evaluated at the scale $k_c$ and pulled outside the integration.

The strategy outlined above is a version of the saddle-point approximation, widely used for one-dimensional variables such as thrust and jet mass. 
It has recently been argued\cite{Aglietti:2025ezs} that this type of approximation may fail to capture relevant contributions in the region it is intended to describe (in our case the low transverse momentum regime) leading to noticeable discrepancies with the ``exact" numerical inversion. 
In the following sections, it will be shown that this is not the case here, and that perfect agreement with the numerical results is recovered, at least for transverse momentum greater than 1-2 GeV.

The same method can be easily extended to the cross section in Eq.~\eqref{eq:tmd_fact}. Introducing the cumulative distribution $d\rho$ up to the cutoff $q_c$, and inserting the Fourier representations of the two PDFs, one obtains:
\begin{align}
    \label{eq:tmd_fact_cumulative}
    &\frac{d \rho (q_c)}{d Q^2 \, d y}
    =
    H(a_S) 
    \int_0^1 d\xi J_1(\xi)
    \sum_q 
    \fourier{F}_{q}\big(\sqrt{\tau} e^{y}, \xi/q_c; Q\big)
    \fourier{F}_{\overline{q}}\big(\sqrt{\tau} e^{-y}, \xi/q_c; Q\big)
\end{align}
where rapidity scales have been canceled out and with $\mu=Q$ and $a_S(Q) \equiv a_S$. The integrand is completely analogous to that of a single TMD distribution. Apart from the substitutions $k_T \to q_T$ and $\mu \to Q$, the only differences are that the Sudakov factor is multiplied by a factor two and the OPE turns into the product of two expansions. For convenience, we introduce the combination
\begin{align}
\label{eq:PDF_comb}
    \mathbf{f}(q_T,\tau,y) =
    \sum_q e_q^2 f_q\big(\sqrt{\tau} e^{y}, q_T \big)
    f_{\overline{q}}\big(\sqrt{\tau} e^{-y}, q_T \big)
\end{align}
that replaces the single PDF appearing at leading order in the OPE. In practical application, a sum over the two colliding hadrons has to be included. The physical cross section is obtained by differentiating Eq.~\eqref{eq:tmd_fact_cumulative} with respect to the cutoff $q_c$.

\subsection{Analytic Resummation up to next-to-leading log}
\label{sec:analytic_res}

At leading-logarithmic (LL) accuracy, the analytic Fourier inversion is straightforward. The TMD distribution in the integrand of Eq.\eqref{eq:Rc_recasted} is approximated by evaluating its dependence on the transverse impact parameter at the value $c_1/k_c$ (or, equivalently, setting $\xi = c_1$). This leaves only the Bessel function to be integrated, and it is well known that this integral evaluates to one. Thus we get:
\begin{align}
    \label{eq:R_LL}
     R_j^{(LL)}(x,k_c) = 
     e^{L_c g_1(\lambda_c)} f_j(x;k_c) 
\end{align}
is agreement with Eq.\eqref{eq:cutoff_scheme}. Its derivative yields the TMD distribution in transverse momentum space at LL:
\begin{align}
\label{eq:F_LL}
    F_j^{(LL)}(x,k_T) = \frac{1}{2\pi k_T^2} e^{L g_1(\lambda)}
    \Big[-h_1(\lambda) f_j(x,k_T) + \frac{d f_j(x,k_T)}{d\log{k_T}}\Big]
\end{align}
where $h_1$ denotes the logarithmic derivative of the Sudakov factor at LL,
\begin{align}
    \label{eq:h1_def}
    h_1(\lambda) = g_1 + \lambda\, g_1'(\lambda) = -\frac{\gamma_K^{[1]}}{2\beta_0} \frac{\lambda}{1-\lambda}
\end{align}
which also coincides with the leading term in the expansion of $H_1$ in Eq.~\eqref{eq:H1}. 
Note that, at low $k_T$, the Landau pole reappears as a branch cut at $\kcut = c_1/\bcut$, with its position shown in Fig.~\ref{fig:bcut}.
Expanding Eq.\eqref{eq:F_LL} in powers of $a_S(\mu)$ correctly reproduces the logarithmic contribution in Eq.\eqref{eq:tmd_NLO_kTspace}, while, as expected for a LL approximation, it does not account for the non-logarithmic term. Note that the last term in  Eq.\eqref{eq:F_LL} corresponds to standard DGLAP evolution, which can be reliably treated in perturbation theory  for $k_T \gtrsim 1$–$2~\mathrm{GeV}$. 
The cross section follows the same analytic structure as Eq.~\eqref{eq:F_LL}:
\begin{align}
    \label{eq:tmd_fact_LL}
    &\frac{d\sigma^{(LL)}}{d Q^2 d y d q_T^2} = 
    H^{[0]} 
    \frac{1}{2\pi q_T^2}
    e^{2 L g_1(\lambda)}
    \Big[
    -2 h_1(\lambda) \mathbf{f}(q_T,\tau,y)
    +
    \frac{d \mathbf{f}(q_T,\tau,y)}{d\log{q}_T}
    \Big]
\end{align}
where $L = \log{(Q/q_T)}$, $\lambda = 2 \beta_0 a_S L$ and $\mathbf{f}$ has been defined in Eq.\eqref{eq:PDF_comb}. This result essentially coincides with the LL expression of Ellis and Veseli~\cite{Ellis:1997ii}. It is free from spurious poles, but it does not reproduce the correct perturbative behavior in the $q_T \to 0$ limit identified long ago by Parisi and Petronzio~\cite{Parisi:1979se}. In fact, the Sudakov resummation in Eq.\eqref{eq:tmd_fact_LL} appears not to break down in this regime, in contrast with expectations. This discrepancy is a direct consequence of the crude nature of the LL approximation, where most of the nontrivial dynamics of transverse momentum is effectively neglected in the integral of Eq.\eqref{eq:Rc_recasted}. To capture these effects, it is necessary to go beyond leading-log accuracy and include the next-to-leading logarithmic contributions.

\bigskip

At next-to-leading-logarithmic (NLL) accuracy, the analytic Fourier inversion becomes non-trivial, as the first correction associated with the function 
$H_1$ must be included explicitly in the integrand of Eq.\eqref{eq:Rc_recasted}:
\begin{align}
    \label{eq:R_NLL_prelim}
     R_j^{(NLL)}(x,k_c) = 
     e^{L_c g_1(\lambda_c) + g_2(\lambda_c)}
     f_j(x;k_c) \,
     \int_0^\infty d \xi J_1(\xi) e^{L_\xi H_1(\lambda_c, L_\xi/L_c)}
     \,.
\end{align}
The integration spans regions where the ratio $L_\xi/L_c$ varies significantly, making the analytic evaluation more involved. The function $H_1$ can be computed explicitly from Eq.\eqref{eq:H1} and the definition of $g_1$ in Eq.\eqref{eq:g1}:
\begin{align}
    \label{eq:H1_explicit}
    &H_1(\lambda_c, r) = 
    \frac{\gamma_K^{[0]}}{2\beta_0}
    \left(
    1+\frac{1}{r\, \lambda_c}\log{\left(
    1-\frac{\lambda_c}{1-\lambda_c} r
    \right)}
    \right)\,.
\end{align}
In the low-transverse momentum region of interest, one might assume that ${L_\xi \ll L_c}$ throughout the integration domain. Under this approximation, ${H_1 = h_1 + \ordof{L_\xi}}$, and the integral can then be solved analytically with relative ease:
\begin{align}
    \label{eq:I_NLL_naive}
    &\int_0^\infty d \xi J_1(\xi) e^{L_\xi H_1(\lambda_c, L_\xi/L_c)}
    \approx
    \int_0^\infty d \xi J_1(\xi) e^{L_\xi h_1(\lambda_c)} 
    =
    \Psi_0\big(h_1(\lambda_c)\big) ,
\end{align}
where
\begin{align}
    \label{eq:Psi0}
    \Psi_0(h_1) = e^{\gamma_E h_1}
    \frac{\Gamma\left(1 + \frac{h_1}{2}\right)}{\Gamma\left(1 - \frac{h_1}{2}\right)}.
\end{align}
Inserting this result back into Eq.\eqref{eq:R_NLL_prelim} produces a consistent logarithmic counting up to NLL when the functions are expanded in powers of $a_S(\mu)$. This coincides with the expression originally obtained by Frixione, Nason, and Ridolfi~\cite{Frixione:1998dw}. Unfortunately, it also exhibits spurious divergences at small and finite transverse momenta, located at
\begin{align}
    \label{eq:FNR_poles}
    &q_T^{\text{\footnotesize nth pole}} = \mu \, \text{exp}\Big\{
    -\frac{1}{a_S(\mu)}\frac{1+n}{2 \beta_0 (1+n) + \gamma_K^{[0]}}
    \Big\},\quad n \geq 0.
\end{align}
at cross section level. Note that the first of these poles already appears at a relatively large scale compared to the Landau singularity, around $1.51$ GeV at the mass of the $Z$ boson. 

The assumption $L_\xi \ll L_c$ breaks down as $\xi \to 0$, where $L_\xi$ becomes large and negative. On the contrary, at large $\xi$ the impact is much less severe, since the logarithm inside $H_1$ provides a strong suppression, well before its non-analyticity point is reached.
Note that the problematic region corresponds to $b_T \ll {c_1}/{k_c}$, which is precisely where highly energetic partonic contributions are expected to spoil the validity of Sudakov resummation at low transverse momentum. 
In fact, Ref.~\cite{Frixione:1998dw} suggested excluding the low-$b_T$ region by introducing a cutoff at $\bmin = c_1/Q$. 
While this prescription successfully removes the spurious divergences, it simultaneously produces contributions that at the cross-section level behave as $q_T^2/Q^2$ when $q_T$ is of order $Q$, thereby spoiling a natural matching with the collinear factorization theorem of Eq.~\eqref{eq:collinear_fact}. 
The problem does not originate from the sharpness of the cut-off: even smoother implementations of the $\bmin$-prescription exhibit the same issues\cite{Cerutti:2025inprep}.
In practice, a smooth matching can still be enforced through external prescriptions such as profiling functions, but this is no longer an intrinsic outcome of the resummation formalism. 
In other words, although these additional terms may superficially resemble higher-twist effects, and one might expect them to be negligible for phenomenology, they are in fact sizable corrections at large transverse momentum, arising solely from the artifacts of the regularization. Moreover, they are multiplied by $a_S \log(Q/q_T)$ contributions, contaminating the logarithmic counting and formally undermining the claimed NLL accuracy through spurious power-suppressed effects. 

Here, we separate the region with $b_T \leq c_1/k_c$ from the region above, where the approximation $L_\xi \ll L_c$ is valid. In the low-$b_T$ region, we do not neglect the contributions; instead, we recognize that the approximation $L_\xi \ll L_c$ breaks down and adopt an alternative treatment. 
The integral in Eq.\eqref{eq:R_NLL_prelim}, restricted to $\xi \geq c_1$ (i.e., $b_T \geq c_1/k_c$), can be easily evaluated analytically up to terms irrelevant at NLL:
\begin{align}
    \label{eq:I_above}
    &\int_{c_1}^\infty d \xi J_1(\xi) e^{L_\xi H_1(\lambda_c, L_\xi/L_c)}
    =
    \notag \\
    &\quad
    \Psi_0\big(h_1(\lambda_c)\big) 
    -\frac{c_1^2}{4}\frac{1}{1+\frac{h_1(\lambda_c)}{2}}{}_1F_2\left(1+\frac{h_1(\lambda_c)}{2};2,2+\frac{h_1(\lambda_c)}{2};-\frac{c_1^2}{4}\right)
    + \dots
\end{align}
where $\Psi_0$ has been defined in Eq.\eqref{eq:Psi0}, while the boundary term is a generalized hypergeometric function ${}_1F_2$. 
The $\dots$ represent terms that are suppressed at NLL.
Note that the cutoff-regularized result of Ref.~\cite{Frixione:1998dw} is obtained from the expression above by replacing $c_1$ with $c_1 \, k_c/Q$. 
The boundary term has several useful properties, collected in Appendix \ref{app:analytic_NLL}. 
Next, we turn to the integral in the low-$\xi$ region, $\xi \leq c_1$, where the previous approximation breaks down. 
Since we cannot approximate the term containing $H_1$, our efforts focuses on the Bessel function $J_1$, which can be expanded in a Taylor series at small $\xi$. The integral then reduces to a series of Laplace transforms of the $H_1$ term, each of which can be evaluated explicitly. The final result reads:
\begin{align}
    \label{eq:I_below}
    &\int_0^{c_1} d \xi J_1(\xi) e^{L_\xi H_1(\lambda_c, L_\xi/L_c)}
    =
    \notag \\
    &\quad
    \frac{c_1^2}{2}
    \sum_{n\geq0} \left(-\frac{c_1^2}{4}\right)^n
    \frac{1}{n!\,(2)_n}
    \frac{1}{c+2(n+1)}
    (A\,z_n)^{-A} e^{A z_n} \Gamma_{1+A}(A \,z_n) 
\end{align}
Here, $\Gamma_{1+A}$ denotes the incomplete Gamma function. We have also introduced
\begin{align}
\label{eq:A_zn_def}
&A = \frac{c \, L_c}{\lambda_c}; \quad
z_n = -\frac{\lambda_c}{h_1(\lambda_c)} \left(c + 2(1 + n)\right),
\end{align}
and $c = \gamma_K^{[0]}/(2 \beta_0)$ for brevity.
The series above converges quickly to the exact value of the integral, with just the first few terms already providing an excellent approximation.
The difficulty in the large-$L_c$ limit becomes apparent: the logarithm cannot be easily separated from the function $h_1$, as $L_c$ enters in two combinations, $L_c/h_1$ and $L_c/\lambda_c$. While $L_c/\lambda_c \approx 1/a_S(\mu)$ is always safely large, $L_c/h_1$ can become small, and eventually vanish, when $\lambda_c = 1$. 

\begin{figure}[t]
    \centering
    \includegraphics[width=1\linewidth]{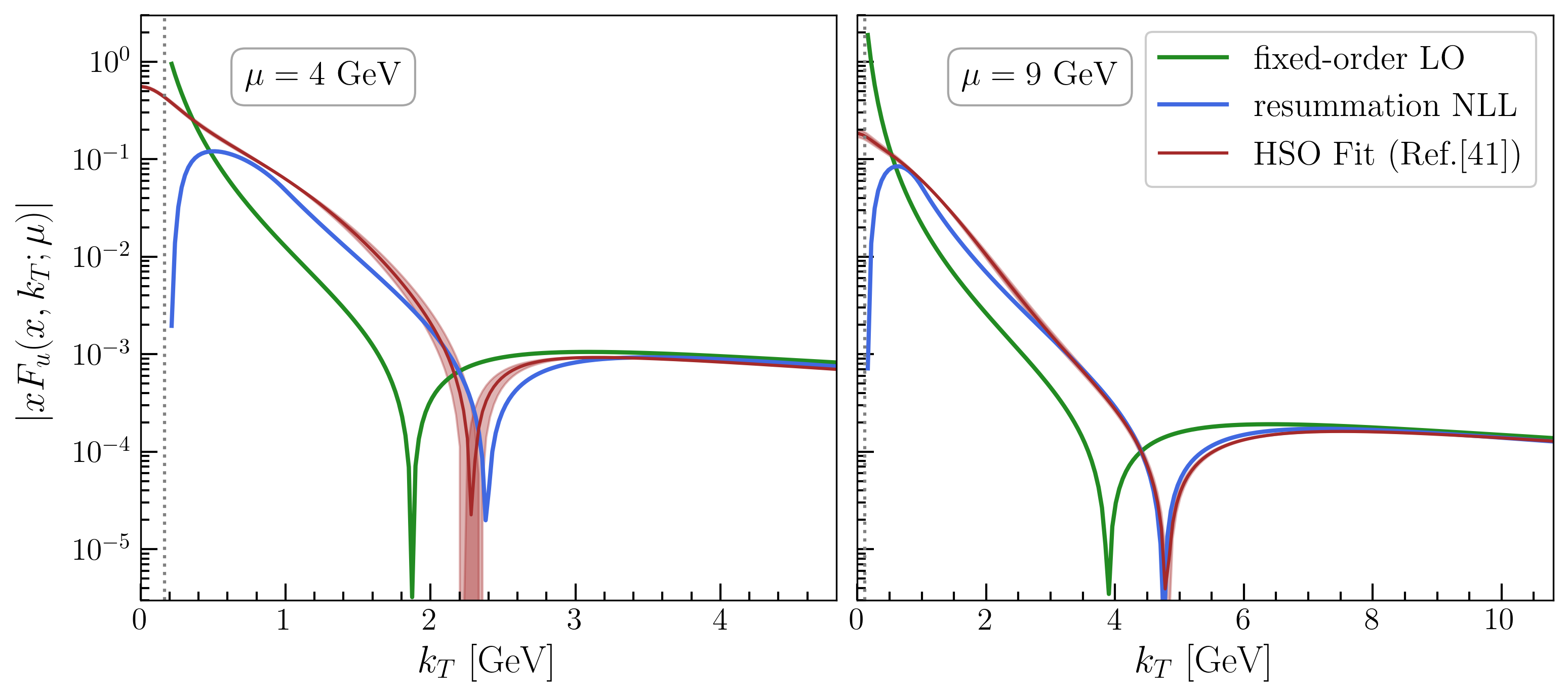}
    \caption{Unpolarized TMD PDF of the up quark in the proton at $\mu=4$ GeV (left) and $\mu=9$ GeV (right), with lightcone momentum fraction $x = 0.1$.
The rapidity scale is fixed to $\zeta=\mu^2$ and the curves are absolute values. Green lines show the fixed-order result of Eq.\eqref{eq:tmd_NLO_kTspace}, while blue lines correspond to the NLL resummation of Eq.\eqref{eq:F_NLL}. 
The red curves correspond to the low-energy fit of Ref.\cite{Aslan:2024nqg}, performed within the ``HSO'' approach on E288 experiment\cite{Ito:1980ev}, with bands indicating the statistical uncertainty. Collinear PDFs are taken from the NLO MMHT set\cite{Harland-Lang:2014zoa}, as in Ref.~\cite{Aslan:2024nqg}. The vertical grey dotted line marks the Landau singularity, located at $k_T \approx 164$ MeV (left) and $k_T \approx 113$ MeV (right).
}
    \label{fig:tmd_match}
\end{figure}

The asymptotic behavior of the series in Eq.~\eqref{eq:I_below} at large and small $k_c/Q$ corresponds, respectively, to the regimes $z_n > 1$ and $z_n < 1$, with $z_n = 1$ marking the poles of the function $\Phi_0$. 
In the OPE region at large $k_c$, the series must cancel the boundary term from Eq.~\eqref{eq:I_above} to ensure a natural matching with the fixed-order. Indeed, the function:
\begin{align}
    \label{eq:Phi}
    &\Phi(\lambda_c,L_c) =
    \frac{c_1^2}{2}
    \sum_{n\geq0} \left(-\frac{c_1^2}{4}\right)^n
    \frac{1}{n!\,(2)_n}
    \frac{1}{c+2(n+1)}
    (A\,z_n)^{-A} e^{A z_n} \Gamma_{1+A}(A \,z_n)
    \notag \\
    &\quad
    -\frac{c_1^2}{4}\frac{1}{1+\frac{h_1(\lambda_c)}{2}}{}_1F_2\left(1+\frac{h_1(\lambda_c)}{2};2,2+\frac{h_1(\lambda_c)}{2};-\frac{c_1^2}{4}\right)
\end{align}
is $\ordof{1/L_c}$ for $z_n > 1$, thus it contributes only at next-to-next-to-leading logarithmic (NNLL) accuracy, confirming that the cancellation is effectively exact at NLL. Moreover, the analytic transparency of this framework makes it possible to quantify such discrepancies systematically. 
In the low transverse momentum region, the Sudakov factor in Eq.\eqref{eq:R_NLL_prelim} is expected to provide a strong suppression. However, the leading exponential term $\sim e^{L_c g_1}$ is cancelled by the asymptotic behavior of the series in Eq.\eqref{eq:I_below}, which instead behaves as $\sim e^{-L_c g_1} k_c^2/Q^2$, thereby reproducing the quadratic suppression first identified by Parisi and Petronzio~\cite{Parisi:1979se} and later confirmed in alternative parton-branching based frameworks~\cite{Monni:2016ktx}.
The explicit derivations underlying these results are collected in Appendix \ref{app:analytic_NLL}. 
Collecting all contributions, the cumulative TMD distribution at NLL reads:
\begin{align}
\label{eq:R_NLL}
R_j^{(NLL)}(x,k_c) =
e^{L_c g_1(\lambda_c) + g_2(\lambda_c)}\,\Psi(\lambda_c,L_c)\,
f_j(x;k_c) 
\end{align}
where, for compactness, we have defined $\Psi(\lambda_c,L_c) = \Psi_0(h_1(\lambda_c)) + \Phi(\lambda_c,L_c)$. The poles in $\Psi_0$ are canceled by those in $\Phi$, leaving no spurious divergences. 
Moreover, in the OPE region where $k_c \approx \mu$, $\Psi$ reduces to $\Psi_0$ up to terms that are at least NNLL, ensuring a natural matching with the fixed order at the claimed NLL accuracy.
Differentiating the expression above yields the TMD distribution:
\begin{align}
    \label{eq:F_NLL}
     &F_j^{(NLL)}(x,k_T) = 
    \frac{1}{2\pi k_T^2}e^{L g_1(\lambda) + g_2(\lambda)} 
    \notag \\
    &\;\times
    \Big\{
    \Big[
    -\Big(h_1(\lambda) + \frac{1}{L} h_2(\lambda) \Big) \Psi(\lambda,L) - 
    \frac{d \Psi(\lambda,L)}{d L}
    \Big] f_j(x;k_T) 
    + \Psi(\lambda,L) \, \frac{d f_j(x;k_T)}{d \log{k_T}}
    \Big\}
\end{align}
where $h_2$ is the NLL correction to the derivative of the Sudakov factor:
\begin{align}
\label{eq:h2_def}
    h_2(\lambda) = g_2'(\lambda) = 
    \Big( \frac{1}{2\beta_0} \frac{\lambda}{1 - \lambda} \Big)^2 
    \Big( \frac{\beta_1}{\beta_0} \gamma_K^{[0]} \log{(1 - \lambda)} - \gamma_K^{[1]} \Big) 
    + \frac{1}{2\beta_0} \frac{\lambda}{1 - \lambda} \gamma_F^{[0]}
\end{align}
also coinciding with the leading term in the expansion of $H_2$ in Eq.\eqref{eq:H2}.
The logarithmic derivative of $\Psi$ consists of two contributions: the term from $\Psi_0$:
\begin{align}
    \label{eq:derPsi0}
    \frac{d\Psi_0\big(h_1(\lambda)\big)}{d L} = 
    \frac{1}{L} \frac{h_1}{1-\lambda} 
    \Psi_0(h_1) \frac{H_{h_1/2} + H_{-h_1/2}}{2}
\end{align}
where $H_n$ is the Harmonic number evaluated at $n$; and the term from $\Phi$: 
\begin{align}
    \label{eq:derPhi}
    &\frac{d\Phi\big(\lambda,L\big)}{d L} = 
    \frac{1}{L} \frac{h_1}{1-\lambda} 
    \frac{c_1^2}{4}\frac{1}{2}\left(-\frac{1}{1+\frac{h_1}{2}}\right)^{2}
    {}_{2}F_{3}\left(1+\frac{h_1}{2},\dots;2,2+\frac{h_1}{2},\dots;-\frac{c_1^2}{4}\right)
    \notag \\
    &\quad+
    \frac{c_1^2}{2}
    \sum_{n\geq0} \left(-\frac{c_1^2}{4}\right)^n
    \frac{1}{n!\,(2)_n}
    \Big[
    1 + \frac{1-z_n}{z_n}(A\,z_n)^{-A} e^{A z_n} \Gamma_{1+A}(A\,z_n)
    \Big]
\end{align}
where the first contribution is the derivative of the boundary term in Eq.\eqref{eq:I_above}, as given by Eq.\eqref{eq:boundary_prop3}. The asymptotic behavior at small and large transverse momentum follows from the features of the cumulative distribution. In particular, the logarithmic derivative of $\Phi$ is NNLL for moderate to large $k_T$, making it a subleading correction in the OPE region and ensuring a natural matching with the fixed order, as confirmed in Fig.\ref{fig:tmd_match}.
At $k_T=\mu$, the discrepancy between the fixed-order and resummed results amounts to about $6.9\%$ at $\mu=4$ GeV, decreasing to $5.3\%$ at $\mu=9$ GeV and $3.3\%$ at the $Z$-boson mass. 

Apart from the region around $k_T \approx \mu$ and the subleading deviations from the fixed-order, it is important to assess how low in $k_T$ the OPE alone provides a reliable estimate of the TMD operator. Ref.\cite{Gonzalez-Hernandez:2022ifv} suggests that this approximation should remain valid at least down to the “node”, i.e. the value where the fixed-order result changes sign, corresponding to the cusp of the green lines in Fig.\ref{fig:tmd_match}. This argument is particularly meaningful at low energies, where the impact of large logarithms is mitigated by the small value of $\mu$.
Indeed, for $\mu \sim 1$–2 GeV, the perturbative and non-perturbative regions tend to shrink and overlap, making the distributions largely non-perturbative. 
However, Fig.~\ref{fig:tmd_match} shows that the resummed curves already deviate from the fixed-order result around the node at low energies, with this behavior setting in from scales of order 1–2 GeV and persisting thereafter.
While this intermediate-$k_T$ region is not directly constrained by low-energy data, it plays a crucial role in determining the quality of predictions at high energy\cite{Cerutti:2025inprep}. Accurately describing the distributions here requires trusting the resummation rather than relying solely on the fixed-order OPE.
In particular, resummation shifts the node to higher $k_T$, effectively pushing the distributions upward and maintaining positivity over a wider range of scales.
Interestingly, this behavior is also seen in an actual extraction of the TMDs from experimental data, as shown by the red curves in Fig.\ref{fig:tmd_match}. These correspond to the fit of Ref.\cite{Aslan:2024nqg}, performed within the Hadron Structure Oriented (HSO) approach on E288 data~\cite{Ito:1980ev}, with bands indicating the statistical uncertainty. Despite including an infrared model as a modification of the NLO fixed-order result, the fit pushes the curves upward, closely aligning them with the blue curves obtained from resummation alone. 
Notably, the uncertainty bands of the fit extend into the intermediate region, while the resummed curves carry no such uncertainty, being entirely dictated by perturbative QCD (PDF uncertainties are not considered here).
Comparison with other standard TMD phenomenological approaches~\cite{Cerutti:2025inprep} confirms this trend, reinforcing that, for $\mu$ above a few GeV, there exists a sizable window from $k_T \gtrsim Q_0 \approx 1$–$2$ GeV up to $k_T \lesssim \mu$, where resummation provides a reliable, essentially model-independent estimate of the TMD distributions, smoothly connecting to the fixed-order OPE at higher $k_T$.
Below this window, an infrared model becomes necessary, as indicated by the departure of the blue curves from the red ones. The blue curves in Fig.~\ref{fig:tmd_match} contain no genuine non-perturbative input beyond the collinear PDFs from the OPE; their behavior at very low scales is entirely dictated by the \verb|LHAPDF| extrapolation\cite{Buckley:2014ana}. In Section\ref{sec:deepIR}, we introduce a physical prescription for analytic continuation into the deep infrared.

\subsection{Comparison with other prescriptions}
\label{sec:compare_prescr}

\begin{figure}
    \centering
    \includegraphics[width=1\linewidth]{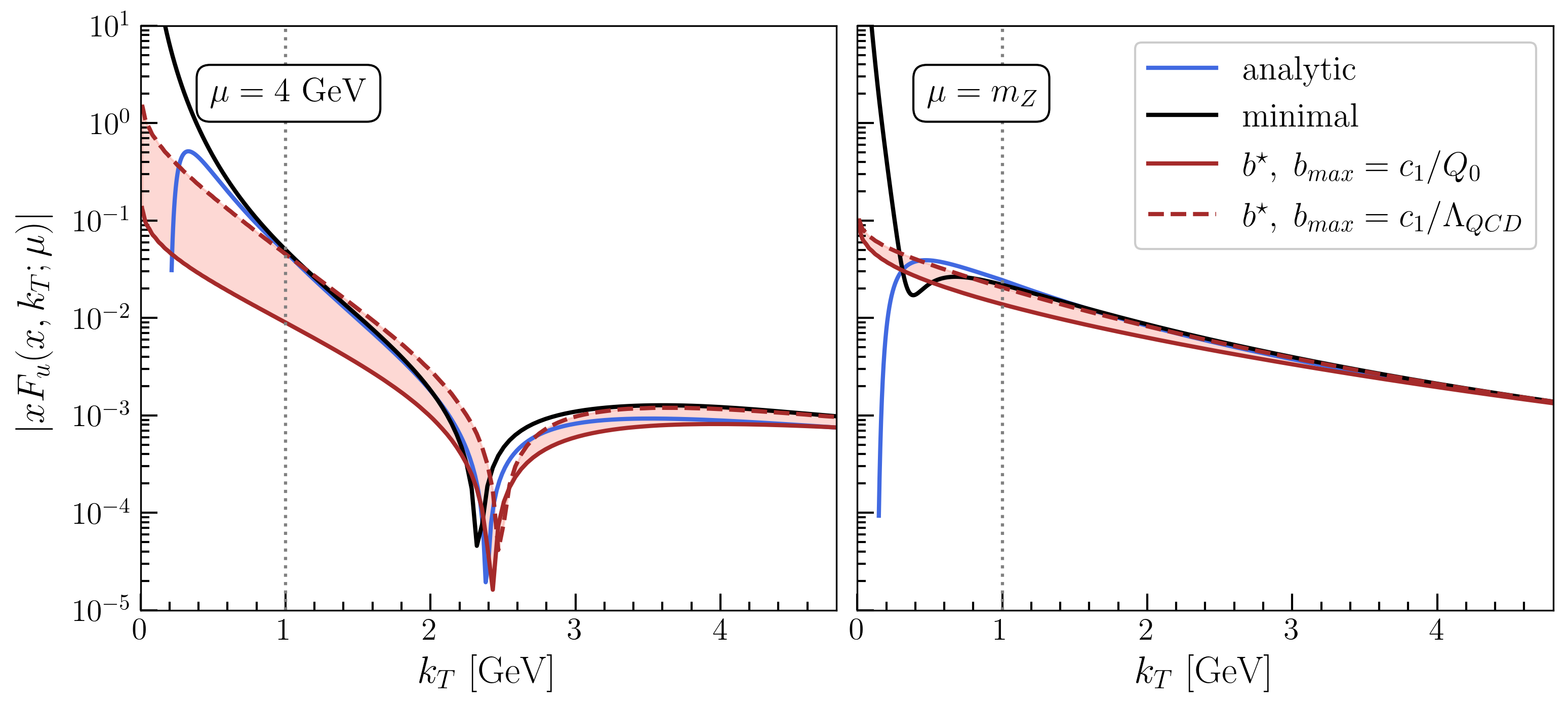}
    \caption{Comparison of different resummation prescriptions at $\mu=4$ GeV (left panel) and at the mass of the  $Z$ boson (right panel) for the up-quark TMD PDF at $x=0.1$ at low transverse momentum. The blue curves show the NLL resummed TMD distributions from Eq.~\eqref{eq:F_NLL}, the black curve corresponds to the minimal prescription, and the red curves represent the $b^\star$ prescription with two choices of $\bmax$: $c_1$ (solid) and $c_1/\Lambda_\text{QCD} \approx 3.38$ GeV$^{-1}$. The vertical grey dotted line indicates the scale $Q_0$, below which the DGLAP evolution is frozen.}
    \label{fig:prescriptions}
\end{figure}

The model-independent nature of our resummation constitutes one of the central results of this work. In contrast, other prescriptions that perform resummation in $b_T$-space necessarily introduce model-dependent non-perturbative corrections. These modifications can distort the distributions in transverse momentum space, particularly in the intermediate region, where pure resummation should suffice to provide a reliable estimate of the TMDs.
The point is that the inverse Fourier transform, when carried out explicitly as a numerical integral, necessarily samples $b_T$ all the way to very large distances, where perturbation theory breaks down and a model input becomes unavoidable. 
To reinforce this point, we compare our formulation with two alternative prescriptions based on the numerical Fourier inversion of the NLL TMD in $b_T$-space of Eq.\eqref{eq:tmd_bT_resummed}:
\begin{itemize}
\item Minimal prescription~\cite{Catani:1996yz,Kulesza:2002rh}: here the integration contour in Eq.\eqref{eq:tmd_fourier} is deformed into the complex plane by exploiting a suitable representation of the Bessel function $J_0$ from the angular integration.
The integral is performed along the real axis up to a cutoff $b_c$, and then continued along two trajectories of the form $b_c - \tau e^{\pm i\phi}$ with $\tau \in [0,\infty)$. 
This procedure requires evolving the collinear PDFs to the scale $c_1/b_c$ on the real axis and  subsequently to complex values. 
In our implementation, we restrict to the real part of $\sim c_1/\tau$, which can reach extremely low values. As a result, the numerical integration along the complex trajectories cannot be carried out consistently unless an infrared model is specified.
\item $b^\star$ prescription~\cite{Collins:1984kg}: here the $b_T$-dependence in the integrand is modified according to
\begin{align}
    \label{eq:bstar}
    b_T^\star = \frac{b_T}{\sqrt{1 + {b_T^2}/{\bmax^2}}}\,,
\end{align}
which saturates at large $b_T$ to the value $\bmax$, as discussed in Section~\ref{sec:bT-space}. The lowest scale at which the collinear PDFs are evaluated thus coincides with $c_1/\bmax$.
Note that this modification alone does not guarantee the convergence of the inverse Fourier transform, since the Bessel function $J_0$ decreases only as $1/\sqrt{b_T}$ at large distances. For this reason, the $b^\star$ prescription is usually supplemented by an additional factor, typically of Gaussian form, which both parametrizes the deep-infrared behavior of the TMD and ensures the convergence of the integral. However, this procedure already extends the model beyond its intended scope of describing the physics at large distances.
In our approach, we avoid this issue by adopting the same strategy used in our analytic framework, since the cumulative TMD distribution $R_j$ does not suffer from convergence problems. Therefore, no extra factor is needed, and the TMD distributions in the $b^\star$ prescription are obtained by differentiating their cumulative counterparts.
\end{itemize}
All of these prescriptions, including the analytic resummation of Eq.\eqref{eq:F_NLL}, require a definition of the collinear PDFs in the deep infrared, at scales below $Q_0 \approx 1$–2 GeV. To enable a meaningful comparison between different approaches, we adopt a simple toy model in which the DGLAP evolution is frozen below $Q_0$, so that the PDFs vary only logarithmically in the deep infrared. This model is purely illustrative and not intended to be physically accurate. Importantly, in the analytic resummation, this extension modifies only the PDFs, while the large logarithms remain fully unregulated, leaving the Landau singularity untouched. 
The result is shown in Fig.~\ref{fig:prescriptions} for both a low-energy and a high-energy case, with $Q_0 = 1$ GeV.
Note that at high energy, all resummation prescriptions essentially coincide for $k_T \gtrsim Q_0$. At low energy, however, the differences become significant. In particular, the analytic resummation (blue curves) remains essentially equivalent to the minimal prescription (black lines), confirming the robustness of the framework developed in this work. 
The comparison with the $b^\star$ curves (red lines) is more instructive. While a relatively large value of $\bmax$ (dashed red lines) roughly agrees with the other prescriptions, the commonly used choice $\bmax = c_1$ (solid red lines) shows a noticeable discrepancy. 
Positioned too far from the Landau cut, it neglects important information from the perturbative region, and should therefore be avoided in extractions, at least for low-energy studies.
However, evaluating collinear PDFs below 1 GeV poses a technical challenge, and unless the deep-infrared DGLAP evolution is explicitly included in the non-perturbative model, this represents the maximal $\bmax$ typically usable in standard approaches. In practice, the discrepancies seen in Fig.~\ref{fig:prescriptions} are compensated by the flexibility of the infrared model, highlighting the limited control standard strategies have over the boundaries between perturbative and non-perturbative regions.

\begin{figure}
    \centering
    \includegraphics[width=.9\linewidth]{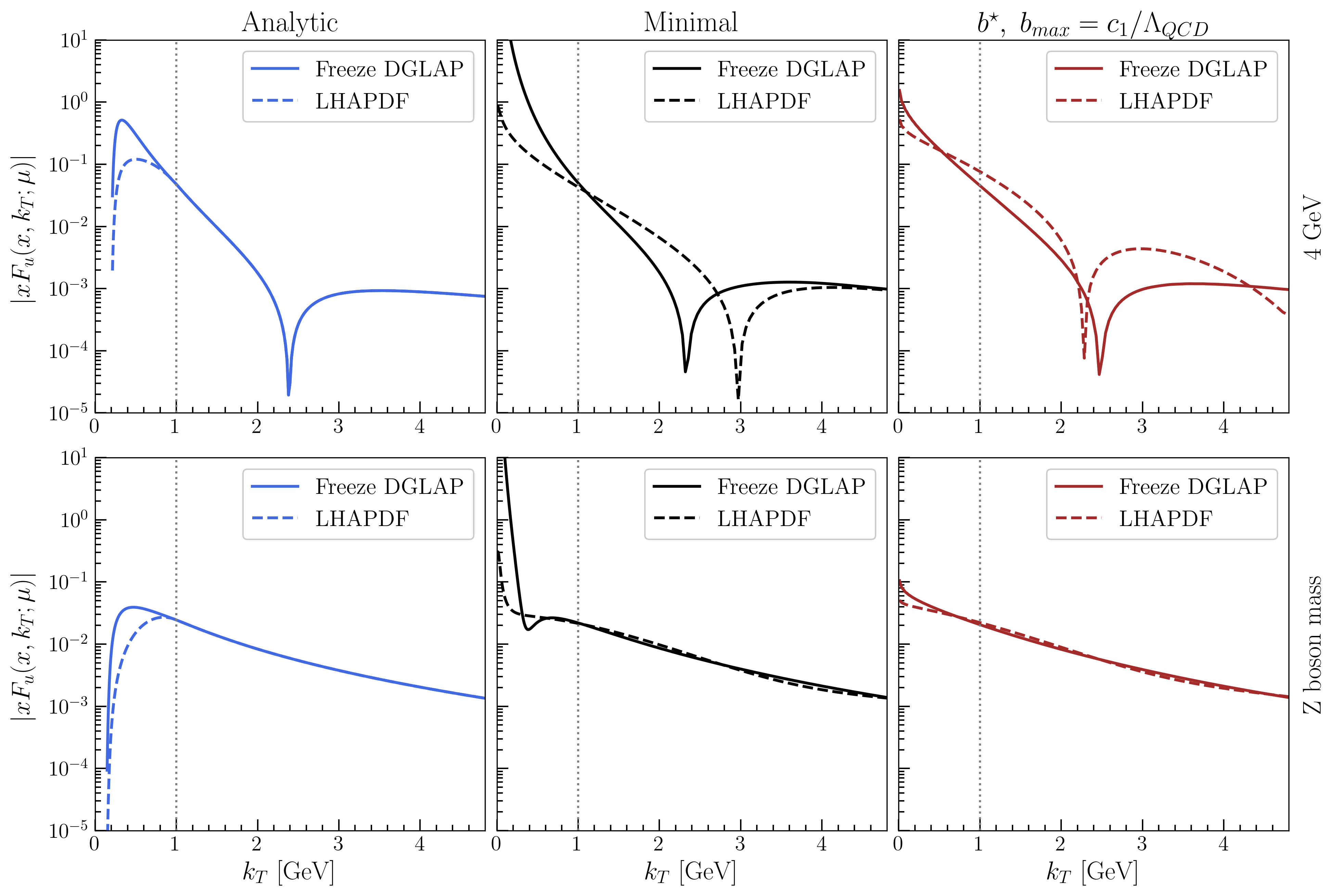}
    \caption{Comparison of two parametrizations for the deep-infrared behavior of collinear PDFs in different resummation prescriptions at $x=0.1$ and two energies: $\mu = 4$ GeV (upper panels) and the $Z$ boson mass (lower panels). The non-perturbative threshold is $Q_0 = 1$ GeV (dotted vertical lines). Solid lines show the toy model with DGLAP evolution frozen below $Q_0$, while dashed lines correspond to PDFs from the \texttt{LHAPDF
    } built-in extrapolation. Columns: left, analytic NLL resummation of Eq.~\eqref{eq:F_NLL}; center, minimal prescription; right, $b^\star$ prescription with $\bmax = c_1/\lambdaQCD$.
}
    \label{fig:nonpert_impact}
\end{figure}

The impact of non-perturbative models is generally difficult to control in approaches based on $b_T$-space resummation, especially at low energies. In Fig.~\ref{fig:nonpert_impact}, we compare the toy model in which the DGLAP evolution is frozen below $Q_0$ (solid lines) with a scenario where the PDFs in the deep infrared are determined solely by the \verb|LHAPDF|      extrapolation (dashed lines), providing an alternative parametrization of extremely low scales\footnote{Except for the case of the $b^\star$ prescription with $\bmax = c_1$, where the two models coincide.}. 
The effects are particularly evident at low energy (upper panels) compared to high energy (lower panels): noticeable discrepancies appear well above $k_T \approx Q_0$ in both the minimal prescription (central column) and the $b^\star$ prescription (rightmost column). In contrast, the analytic framework (leftmost column) remains unaffected by these choices beyond $Q_0$, ensuring that only perturbative QCD governs $k_T \gtrsim Q_0$ while the non-perturbative modeling is confined below this threshold. 
In conclusion, at high energies the choice of resummation prescription has little impact on the results, except for extremely low $k_T$, which typically affects only one or two bins in high-energy data. By contrast, at low energies most data lie below $Q_0$, making it very easy to lose control over the region where non-perturbative models influence the distributions. The fully analytic, $k_T$-space approach devised in this work provides a robust alternative, allowing the predictive power of perturbative QCD to be fully exploited even at low energies, while ensuring maximal control over the interface between perturbative and non-perturbative regimes.

\subsection{Test against experimental data at high energy}
\label{sec:test_data}

The cross section for Drell–Yan scattering, resummed analytically in $q_T$-space at next-to-leading logarithmic (NLL) accuracy, follows the same strategy used to derive the TMD distribution in Eq.\eqref{eq:F_NLL}, with the substitutions already applied to obtain the cross section at leading-log (LL) in Eq.\eqref{eq:tmd_fact_LL}.
These substitutions require a slightly more careful treatment in this case, as we need to adapt the function $\Phi$ of Eq.\eqref{eq:Phi} (and its derivative) to the cross section. This amounts to replacing $z_n \to z'_n = {\lambda}/{h_1}\,(c + n + 1)$ and $A \to A' = 2A$ in Eq.\eqref{eq:A_zn_def}. 
The final result has the same analytic structure of Eq.\eqref{eq:F_NLL}:
\begin{align}
    \label{eq:tmd_fact_NLL}
    &\frac{d\sigma^{(NLL)}}{d Q^2 d q_T^2 d y} = 
    H^{[0]} 
    \frac{1}{2\pi q_T^2}e^{2 \big(L g_1(\lambda) + g_2(\lambda)\big)} 
    \notag \\
    &\,\times
    \Big\{
    \Big[
    -2 \Big(h_1(\lambda) + \frac{1}{L} h_2(\lambda) \Big) \Psi'(\lambda,L) - 
    \frac{d \Psi'(\lambda,L)}{d L}
    \Big] \mathbf{f}(q_T;\tau,y)
    + 
    \Psi'(\lambda,L) \, \frac{d \mathbf{f}(q_T;\tau,y)}{d \log{q_T}}
    \Big\}
\end{align}
Here, $L = \log(Q/q_T)$ and $\lambda = 2 \beta_0 a_S L$. The prime on the function $\Psi'$ indicates that it is obtained from $\Phi$ after the substitutions described above, while $\mathbf{f}$ denotes the combination of collinear PDFs defined in Eq.~\eqref{eq:PDF_comb}.
Following the arguments discussed in the previous section, this expression provides a reliable estimate of the Drell-Yan cross section for $q_T \gtrsim Q_0$, where $Q_0$ marks the onset of the non-perturbative regime. It naturally reproduces the asymptotic behavior predicted by the collinear factorization in Eq.~\eqref{eq:collinear_fact} at low transverse momentum, without the need for any additional profiling functions.

\begin{figure}
    \centering
    \includegraphics[width=1\linewidth]{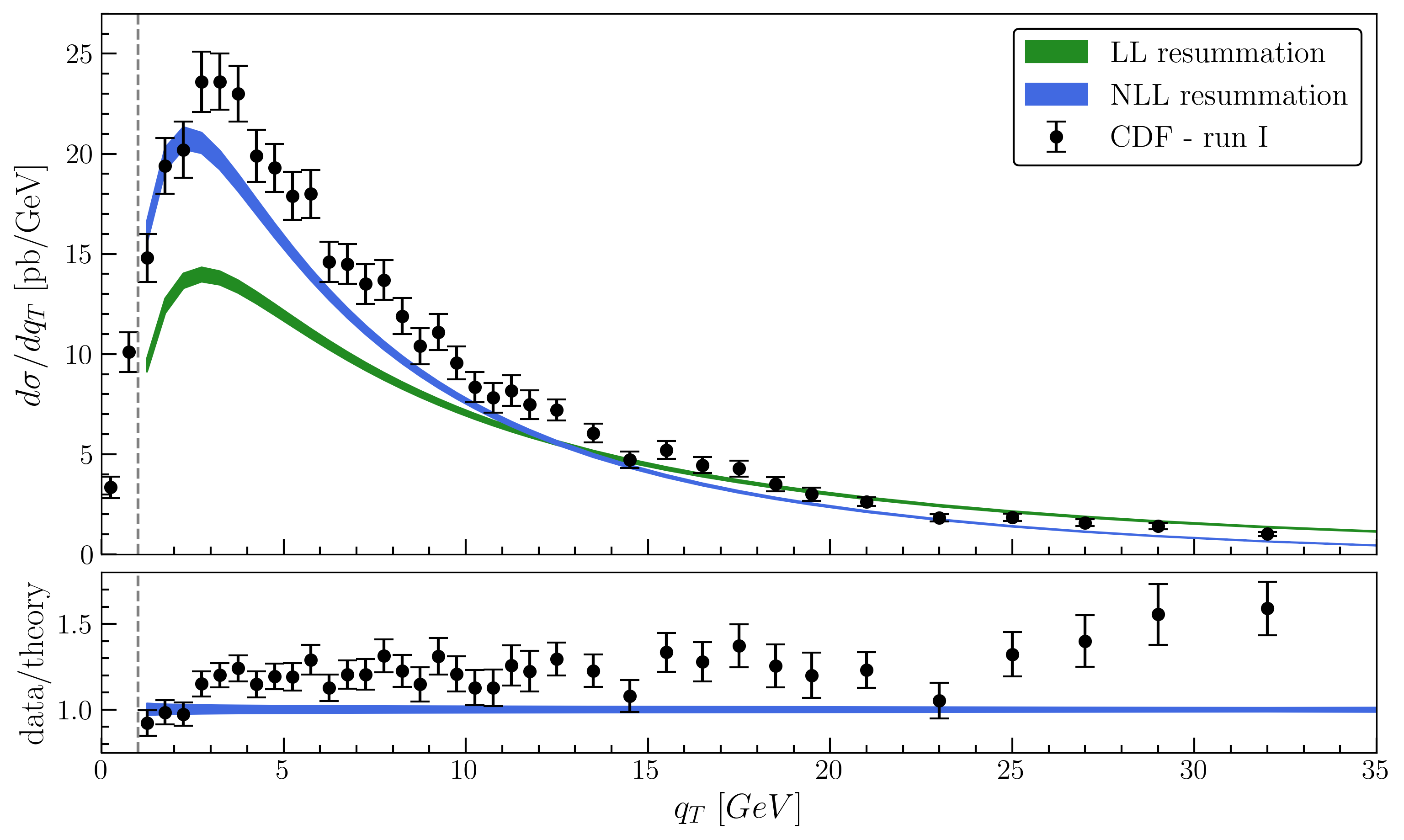}
    \caption{Comparison of the analytic resummation of the cross section at leading-log from Eq.\eqref{eq:tmd_fact_LL} (green band) and next-to-leading-log (blue band) from Eq.\eqref{eq:tmd_fact_NLL} with $Z$ boson production data from the CDF Run I data set\cite{CDF:1999bpw}. Error bars on data points represent statistical uncertainties only; systematic and normalization uncertainties are not included. The theory bands correspond to the 68\% PDF uncertainty of the NLO MMHT set\cite{Harland-Lang:2014zoa}. The lower panel shows data and theory normalized to the central theory prediction at NLL. The grey vertical dotted line indicates the threshold $Q_0 = 1$ GeV, below which non-perturbative effects are expected. No fit to data has been performed: the curves represent pure theory predictions.}
    \label{fig:CDF_data}
\end{figure}

This expression, together with the leading-log cross section of Eq.\eqref{eq:tmd_fact_LL}, can be tested against high-energy experimental data, where the resummation window $Q_0 \lesssim q_T \ll Q$ is sufficiently large to assess the validity of the resummation framework developed here. 
In Fig.\ref{fig:CDF_data}, we compare our predictions with the CDF Run I data set\cite{CDF:1999bpw} for $q_T \leq 35$ GeV. In fact, beyond $q_T \approx$ 25-30 GeV, the TMD factorization gradually deteriorates as shown in the lower panel of  Fig.\ref{fig:CDF_data}, and the interplay with collinear factorization needs to be properly accounted for. 
The explicit matching will be implemented in future work.
On the other hand, for $q_T \lesssim Q_0$, here set to 1 GeV and marked by the grey dotted vertical line in Fig.~\ref{fig:CDF_data}, non-perturbative effects become unavoidable. A consistent strategy to incorporate them will be discussed in the next Section\ref{sec:deepIR}.
The figure clearly illustrates the significant improvement obtained when going from LL to NLL accuracy. It is important to emphasize that no fit is involved in this comparison: the curves represent genuine theory predictions for the cross section, derived from the LL expression of Eq.\eqref{eq:tmd_fact_LL} (green band) and the NLL expression of Eq.\eqref{eq:tmd_fact_NLL} (blue band). The bands reflect solely the uncertainty of the collinear PDFs, taken from the NLO MMHT set\cite{Harland-Lang:2014zoa} at the 68\% confidence level. At large $q_T$, where matching with collinear factorization becomes necessary, this uncertainty is about 2–3\%, comparable to the NNLL discrepancy discussed in the previous Section. Overall, the deviation of the NLL prediction from the data amounts to a normalization offset of about 10–15\% for $q_T \lesssim 25$ GeV, which can be partly accounted for by the systematic uncertainty not included in the present comparison. The eventual remaining difference can be addressed by further improving the description by extending the resummation beyond NLL, as suggested by the trend observed in the plot.

In contrast to the perspective of Ref.~\cite{Aslan:2024nqg,Rogers:2024cci}, where high-energy data are used to discriminate among non-perturbative models tuned at low energies, and to the standard global-fit strategy\cite{Bacchetta:2022awv,Moos:2023yfa,Bacchetta:2024qre,Moos:2025sal,Bacchetta:2025ara}, where high- and low-energy data are simultaneously fitted with highly flexible parametrizations, the framework developed here follows a different path. 
Within our fully analytic approach, high-energy data require minimal phenomenological modeling: the only effort lies in pushing the resummation to higher logarithmic accuracy, thereby maximizing theoretical control and predictive power while clearly isolating the role of non-perturbative dynamics.
Efforts to determine appropriate infrared models remain necessary at low energies, where most data lie below the threshold $Q_0$.

In global fits, the description of high-energy data is particularly sensitive to the choice of the $b^\star$ prescription, as variations in it can induce sizable effects not only at very small $q_T$, but also in the intermediate region\cite{Cerutti:2025inprep}. 
Once the $b^\star$ prescription is fixed, increasing the logarithmic accuracy of the perturbative calculation systematically improves the predictions, with the impact of the non-perturbative model remaining confined to low transverse momentum. 
However, the precise onset of the non-perturbative regime can only be determined a posteriori, after the fit. Moreover, changing the $b^\star$ prescription effectively reshuffles the boundaries between the perturbative and non-perturbative regions. By exploiting both low- and high-energy data, the global fit partially compensates for these distortions and still provides good agreement with the measurements, while simultaneously absorbing a variety of poorly controlled effects, potentially blurring the identification of genuine non-perturbative dynamics. In this way, it soon becomes difficult to disentangle whether improvements at low-to-moderate transverse momentum are genuinely due to higher-order resummation or rather induced by phenomenological modeling.

The change of perspective proposed here not only fully exploits the predictive power of perturbative QCD, but also gives a deeper meaning to the non-perturbative models themselves. Rather than highly flexible parametrizations devised to reproduce experimental data, they become genuine probes of the deep infrared regime of QCD, a topic we address in the next Section.

\section{Analytic continuation into the Deep Infrared}
\label{sec:deepIR}

In the previous Sections, we introduced the scale $Q_0$, which marks the onset of non-perturbative dynamics, and the door of the deep infrared regime. Its precise value cannot be predicted within perturbation theory, but it is expected to lie in the few-GeV range, typically around $1$–$2$ GeV. Possible physical interpretations include its identification with the proton mass or with the characteristic scale of chiral symmetry breaking. 
Below this threshold, perturbative QCD is no longer reliable and, strictly speaking, all resummed expressions derived in the previous Sections cease to be trustworthy. In this region, the TMD distributions are not constrained by perturbation theory: they can, in principle, take any functional form as long as they continuously match the perturbative resummed result at $k_T \approx Q_0$. The same consideration applies to the cross section itself.
This ``agnostic” perspective has led to the adoption of increasingly flexible parametrizations, culminating in the use of neural networks\cite{Bacchetta:2025ara}, which aim to reconstruct the shape of TMD operators deep in the infrared region, without imposing additional theoretical constraints. 
In this way, one may capture the overall trend of the distributions, but not their genuine functional form. While the ultimate goal is to achieve a faithful tomography of hadrons, this strategy inevitably sacrifices physical interpretability together with analyticity. The specific functional shape of the parton distributions (whether they follow, for instance, a Gaussian profile or a power-law behavior) becomes elusive, and only broad qualitative features can be inferred. The very notion of a simple analytic shape is lost, and with it the possibility of a physically transparent interpretation.

Here, we suggest a different perspective. Since the formalism of the previous sections provides the analytic structure of the distributions above $Q_0$, we propose to extend it below the threshold by analytic continuation. This means not adopting a generic parametrization, but rather enforcing a specific functional form, consistent with the perturbative structure, that is properly modified to remain valid in the deep infrared. 
A consistent analytic continuation of the resummed expressions below $Q_0$ hinges on a careful and controlled extension of at least two crucial ingredients, both of which determine the behavior of QCD in the deep infrared:
\begin{itemize}
\item The strong coupling $\alpha_S$ below $Q_0$, or equivalently the QCD Beta function.
\item The collinear PDFs below $Q_0$, or equivalently the DGLAP splitting kernels.
\end{itemize}
The decisive feature enabling this perspective is that both the strong coupling and the PDFs enter the resummed TMD factorization theorem at the scale $q_T$, rather than the hard scale $Q$ as in collinear factorization. This scale shift is pivotal: in low-energy experiments, $q_T$ can probe deeply into the infrared, well below 1–2 GeV, making the analytic continuation into this region both meaningful and necessary. 

\begin{figure}
    \centering
    \includegraphics[width=0.7\linewidth]{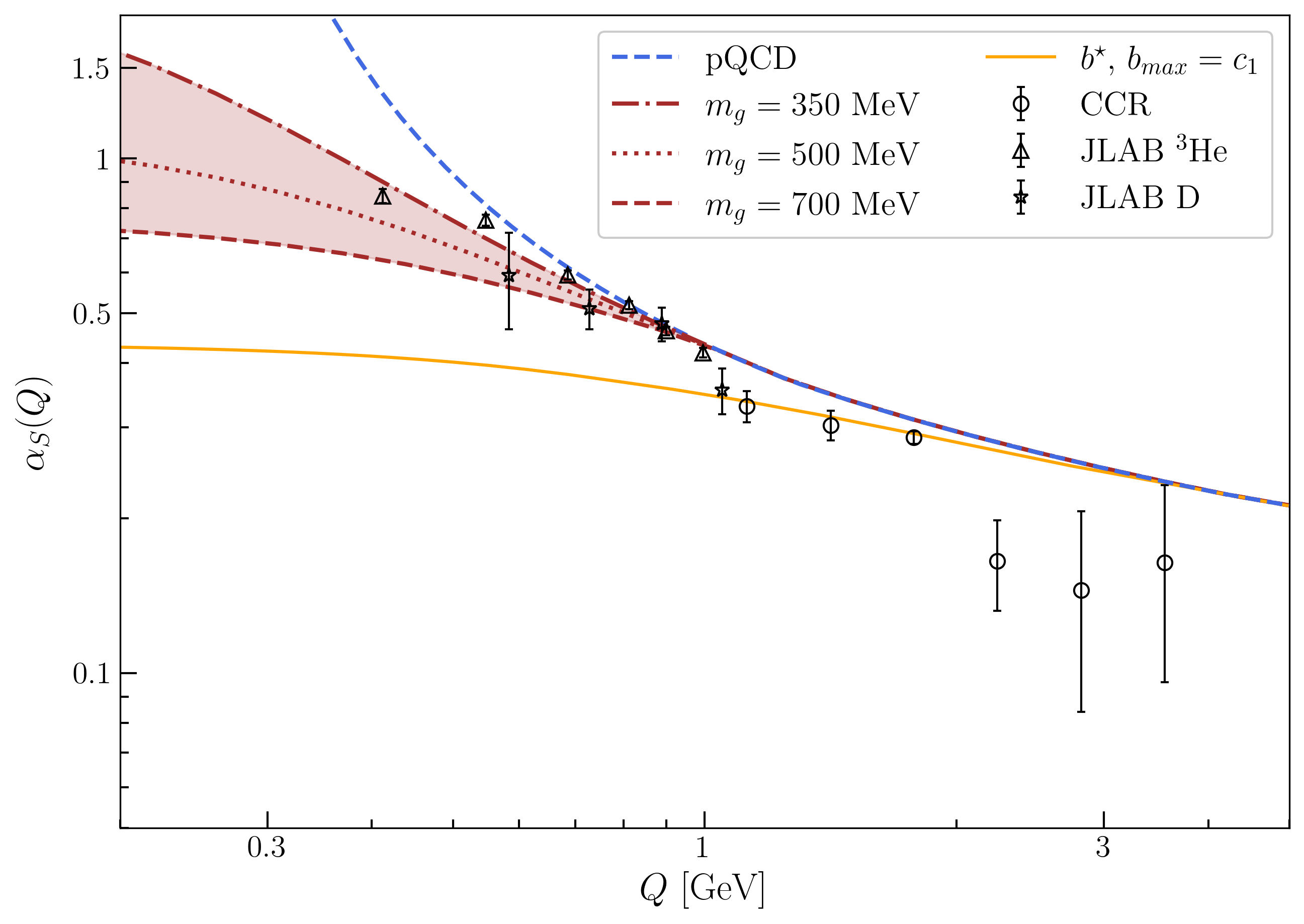}
    \caption{Comparison of different parametrizations of the strong coupling in the deep infrared with experimental data. Star and triangle markers correspond to JLab measurements\cite{Deur:2004ti} from inclusive polarized electron scattering off deuterons and ${}^3$He, respectively, while circle markers show neutrino data from the CCFR Collaboration\cite{Kim:1998kia}. Error bars include only statistical uncertainties; systematic uncertainties are not considered. The blue dashed curve represents the perturbative NLL solution of the renormalization group evolution of $a_S$, cf. Eq.\eqref{eq:aSmub_RG}. The red curves correspond to the model of Eq.\eqref{eq:alphaS_model} for three representative values of $m_g$ and with $Q_0 = 1$ GeV. The orange curve shows the result of the $b^\star$ prescription with the standard choice $\bmax = c_1$.}
    \label{fig:alphaS}
\end{figure}

In the cross sections of Eq.\eqref{eq:tmd_fact_LL} and Eq.\eqref{eq:tmd_fact_NLL}, the strong coupling appears both explicitly at the hard scale $Q$ through $\lambda$, and implicitly at the transverse scale $q_T$ via the logarithms $\log(Q/q_T)$, which are effectively traded for $a_S(q_T)$ via the renormalization group. 
At very small $q_T$, the logarithms become large and positive, eventually encountering the Landau singularity at a few hundred MeV.
To extend $\alpha_S$ below $Q_0$, we modify the $q_T$ dependence inside the logarithms themselves, leaving them unchanged above the threshold but altering their behavior smoothly at low transverse momentum. 
A possible parametrization is:
\begin{align}
\label{eq:alphaS_model}
\qTeff =
\begin{cases}
q_T & \text{if } q_T \ge Q_0,\\[2mm]
Q_0 \sqrt{\dfrac{q_T^2 + m_g^2}{Q_0^2 + m_g^2}}+ & \text{if } q_T < Q_0.
\end{cases}
\end{align}
where we have introduced the parameter $m_g$, interpreted as an effective gluon mass of order $\Lambda_{\mathrm{QCD}}$. This modification saturates the transverse momentum in the deep infrared around $m_g$, which in turn leads to a saturation of the strong coupling at low scales as conjectured by several nonperturbative theoretical approaches\cite{Mattingly:1993ej,Deur:2016tte}. 
Note that this perspective implicitly assumes that the long-distance behavior of the Collins–Soper kernel, perfectly disentangled and identifiable in $b_T$-space but lost in the convolutions in transverse momentum space, is entirely determined by the deep-infrared behavior of the strong coupling.
Also, this modification resembles the $b^\star$ prescription discussed in the previous Section, but it is conceptually and practically different. First, in its intent: Eq.~\eqref{eq:alphaS_model} is not merely a prescription to avoid the Landau singularity, but an explicit parametrization of the strong coupling in the deep infrared, controlled by the physical parameter $m_g$ with a clear interpretation. Second, in its implementation: here the threshold scale $Q_0$, at which non-perturbative effects appear, is separated from the scale $m_g$ where $\alpha_S$ saturates, whereas in most $b^\star$ applications one sets $\bmax = c_1 / Q_0$, effectively making the two scales coincide.
This distinction between the ``transition" scale $Q_0$ from the ``saturation" scale $m_g$ has also been proposed in the approach of Refs.~\cite{Gonzalez-Hernandez:2022ifv,Gonzalez-Hernandez:2023iso}, although without a direct connection to the strong coupling.
It is notoriously difficult to obtain an experimental determination of $\alpha_S$ at such low energy scales. One possible strategy is to extract it from the Bjorken sum rule\cite{Deur:2005cf,Deur:2008rf}, thereby defining effective strong couplings\cite{Grunberg:1980ja,Brodsky:1994eh} that absorb both nonperturbative effects and higher-order perturbative contributions. As an illustration, in Fig.\ref{fig:alphaS} we compare the strong coupling resulting from our model, Eq.\eqref{eq:alphaS_model}, with that obtained through the common $b^\star$ prescription with $\bmax = c_1$, confronting both with the JLab measurements\cite{Deur:2004ti} and with data from the CCFR Collaboration\cite{Kim:1998kia}, without considering the systematic uncertainties.
Compared to the $b^\star$ parametrization, the deep-infrared model of $a_S$ proposed in Eq.~\eqref{eq:alphaS_model} provides greater flexibility and exhibits a better agreement with experimental data below 1 GeV. For instance, an effective gluon mass of about 500 MeV naturally yields a saturated value of $a_S \approx 1$ at zero energy. 
Moreover, the model is explicitly constructed so as to leave unaltered the perturbative solution of the renormalization-group evolution at scales larger than $Q_0$, while in the $b^\star$ prescription the transition between perturbative and nonperturbative regimes is less controlled. The precise values of the relevant nonperturbative parameters, such as $Q_0$ and $m_g$, as well as the functional form of the freezing mechanism, must ultimately be determined through low-energy TMD phenomenology.

\begin{figure}
    \centering
    \includegraphics[width=0.7\linewidth]{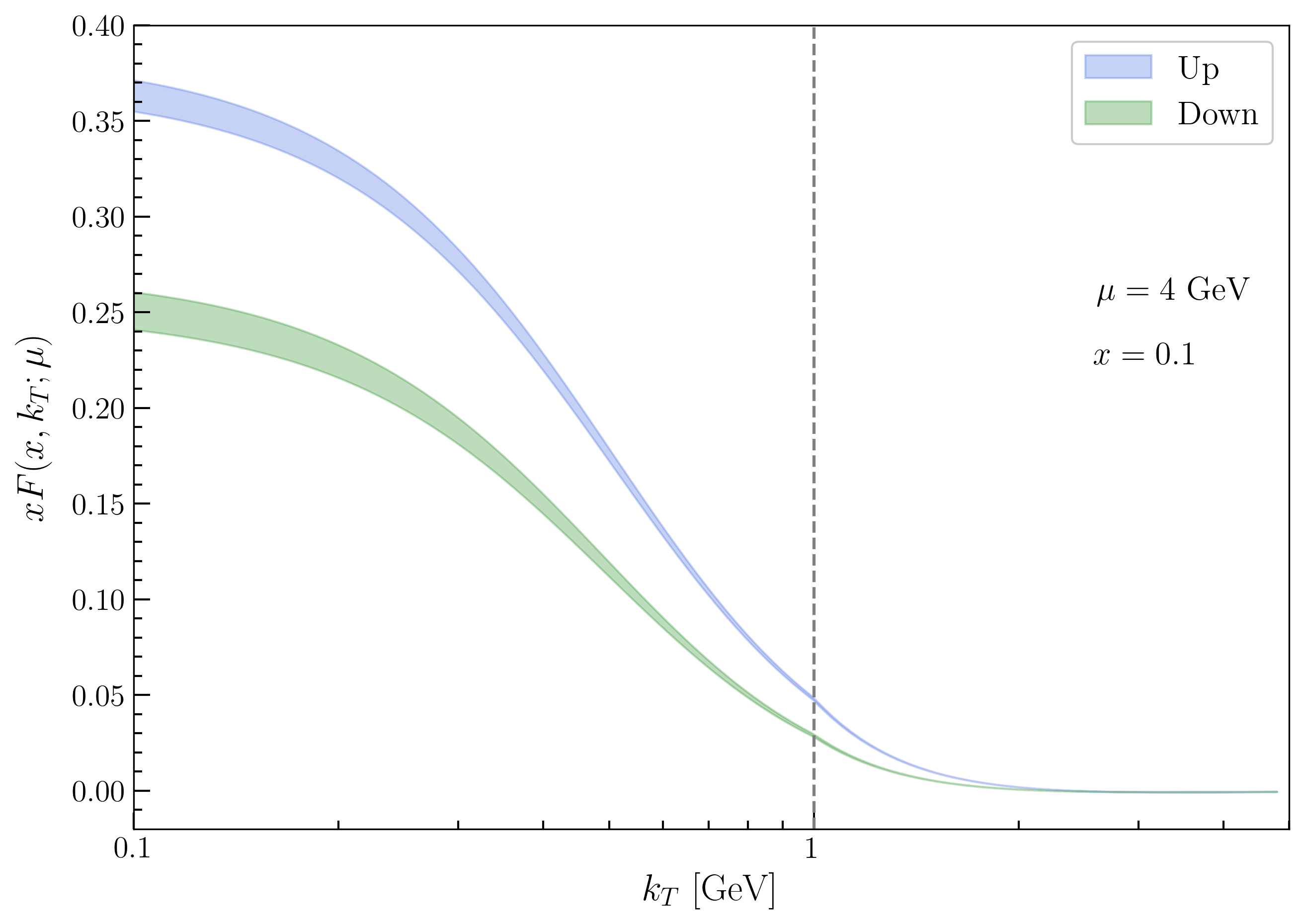}
    \caption{Unpolarized TMD PDFs for up (blue) and down (green) quarks in the proton, at $x=0.1$ and $\mu = 4$ GeV, with rapidity scale $\zeta=\mu^2$. The bands indicate the 68\% C.L. uncertainties of the collinear PDFs, taken from the NLO MMHT set\cite{Harland-Lang:2014zoa}. The perturbative content is evaluated at NLL accuracy according to Eq.\eqref{eq:F_NLL}. The vertical gray dashed line marks $Q_0=1$ GeV: above this threshold the result follows purely from Eq.\eqref{eq:F_NLL}, while below non-perturbative effects are included, namely the infrared model for $\alpha_S$ of Eq.~\eqref{eq:alphaS_model} and the frozen–DGLAP model for collinear PDFs.}
    \label{fig:collpdfs}
\end{figure}

The second source of non-perturbative effects originates from the collinear PDFs. In collinear factorization, these objects are probed only at sufficiently large energy scales, typically above 1–2 GeV, where the factorization theorem is valid. In contrast, within TMD factorization the PDFs naturally enter at the scale $q_T$, which can easily fall below the non-perturbative threshold $Q_0$. This makes their extension into the deep infrared unavoidable. It is important to stress, however, that as QCD operators the PDFs are rigorously defined at any scale. The usual practice of introducing them only above a low input scale $Q_0 \sim 1$ GeV is a pragmatic choice, tied to the regime where collinear factorization applies, rather than a fundamental limitation.
Parametrizing the PDFs in the deep infrared is, in practice, equivalent to specifying the behavior of the DGLAP splitting kernels at very low scales. In the previous Section we illustrated this point with a simple toy model, in which DGLAP evolution was assumed to freeze below $Q_0$, leading to the deviations observed in Fig.\ref{fig:nonpert_impact}.
Note that the standard approach adopted in TMD extractions can be viewed as a particular case of this perspective, in which the PDFs in the deep infrared are effectively frozen at the scale $c_1/\bmax$ and multiplied by the non-perturbative factors $F_{NP}$ typical of the conventional strategy\footnote{This parametrization does not guarantee continuity in the energy-derivative of the PDFs at the perturbative/non-perturbative interface.}.
Although the frozen–DGLAP model has no ambition of being physically accurate, it already illustrates an important feature of this framework: the infrared behavior of TMDs naturally acquires a flavor dependence. This aspect, which has only recently started to be systematically addressed in phenomenological analyses\cite{Bacchetta:2024qre,Moos:2025sal}, is clearly visible in Fig.~\ref{fig:collpdfs}, where the up and down-quark TMD PDFs inside proton exhibit markedly different patterns.
The figure illustrates a toy realization of the perspective advocated in this work. Above the infrared threshold $Q_0$, the distributions are entirely determined by perturbative QCD at NLL accuracy via Eq.~\eqref{eq:F_NLL}. Below this scale, they are analytically continued by modifying the behavior of both the strong coupling and the PDFs in the deep infrared. We emphasize once more that the specific functional form of this continuation must ultimately be determined by TMD phenomenology, in particular from low-energy experimental data.
In this perspective, collinear PDFs are a cornerstone: together with $\alpha_S$, they provide the foundation upon which TMD distributions, and consequently the low-$q_T$ cross sections, are constructed. This framework opens the possibility of a consistent determination of PDFs directly from transverse-momentum–dependent data. Input functions can be defined at the scale $k_T = Q_0$, evolved perturbatively to higher energies, and, through a suitable non-perturbative continuation, evolved backwards into the deep infrared. 
The TMD phenomenology itself thus becomes an invaluable tool to probe the foundations of QCD precisely where they are least constrained, at extremely low energy scales.

\bigskip

The approach suggested in this work, where resummation is performed directly in transverse momentum space, has many advantages widely discussed in the previous Sections, but also some intrinsic limitations. For example, the neat factorization structure of $b_T$-space is no longer manifest as the two TMDs that enter Eq.\eqref{eq:tmd_fact} become deeply intertwined in Eq.\eqref{eq:tmd_fact_NLL}. We conclude this Section by suggesting a possible strategy to reconcile the present framework with the more traditional $b_T$-space formalism.
The proposed strategy develops in two steps:
\begin{enumerate}
\item \emph{Infrared modeling}. The first step is devoted to identifying suitable parametrizations for the strong coupling, via a modified $q_T$-dependence of the logarithms $\log(Q/q_T)$ for $q_T \leq Q_0$, and for the collinear PDFs, via infrared extensions of the DGLAP splitting kernels. This analysis is performed directly in transverse-momentum space, by analytically continuing the resummed cross section below $Q_0$. In this stage, low-energy data play a central role in constraining the parametrizations.
\item  \emph{TMD extraction}. Once the infrared behavior of $\alpha_S$ and the PDFs has been inferred at the level of cross sections and tested against data across a wide energy range, the successful functional forms can be re-implemented in $b_T$-space. Here the inverse Fourier transform is carried out numerically, avoiding any cross-contamination between regions since the functional forms have already been validated. 
For example, one could deform the logarithms in $b_T$-space inspired by the model of Eq.\eqref{eq:alphaS_model}, inducing a saturation of $\alpha_S$ at the scale $\approx c_1/m_g$; to be compared with the significantly smaller value of the conventional $b^\star$ choice $\bmax = c_1$. Any ambiguities in the definition of PDFs below 1 GeV would be overcome, since the backward non-perturbative evolution has already been tested and validated in the first step. In this way, the fit can be repeated and the TMDs extracted explicitly, taking advantage of their disentangled structure in $b_T$-space.
\end{enumerate}
The core of this strategy lies in the analytic resummation and its controlled continuation into the deep infrared. Nevertheless, the steps outlined above illustrate both the versatility and the full potential of the perspective suggested here. By adopting this approach, all the difficulties highlighted in Section~\ref{sec:bT-space} are circumvented, allowing for a more rigorous and transparent control of non-perturbative dynamics and the connection with fundamental properties of strong interactions.

\section{Conclusions}

Observables sensitive to transverse momentum have historically been studied in the auxiliary impact-parameter space, where factorization is manifest and the corresponding evolution equations admit simple solutions. Reaching the same level of analytic control directly in transverse-momentum space, however, has long remained a challenge. Available formulations typically relied on infinite series of convolutions which, although analytic, did not provide a closed form. 
The origin of these difficulties can be traced back to the vectorial nature of transverse momentum: a small net $q_T$ can still result from the cancellation of large partonic transverse recoils. This spoils the transparent Sudakov-type resummation and, if not treated carefully, may generate dangerous spurious singularities at low $q_T$.
In this work, we have overcome this limitations by obtaining a closed analytic expression for the resummed cross section up to next-to-leading logarithmic accuracy. Our formalism reproduces the known results at leading logarithmic accuracy\cite{Ellis:1997sc} and correctly recovers the expected quadratic perturbative behavior in the vanishing transverse-momentum limit\cite{Parisi:1979se}. 
This analytic control makes it possible to exploit perturbative QCD to its full extent for transverse momenta above $\sim$1–2 GeV and hence, most importantly, to unambiguously identify the threshold scale $Q_0$ that marks the interface between perturbative and non-perturbative dynamics. The precise value of this scale will ultimately have to be determined phenomenologically, but it is reasonable to expect it to lie in the range 1–2 GeV, possibly connected with fundamental hadronic scales such as the proton mass or the scale of chiral symmetry breaking. 

At high energies, the vast majority of experimental measurements lie safely within the perturbative domain, with only a handful of bins probing transverse momenta below $Q_0$. In this regime, the analytic resummation alone is expected to account for the data, with non-perturbative effects confined to the very first bins. Accordingly, within our framework, the phenomenological description of high-energy measurements depends almost entirely on the theoretical effort of pushing the resummation to higher logarithmic accuracy. This expectation is confirmed by the comparison with CDF Run I data in Fig.\ref{fig:CDF_data}, where the transition from LL to NLL accuracy yields a clear and significant improvement.
A systematic application of this formalism to high energy data will require the extension of the framework beyond NLL accuracy. Such an upgrade will be solidly founded on the result of this work, as the conceptual challenge has already been overcome at NLL, where the issue of spurious poles first arises and is resolved.

At low energies, most experimental measurements lie below $Q_0$, making the inclusion of non-perturbative corrections unavoidable. The framework presented here allows for a consistent and well-defined analytic continuation of the resummed expressions into the deep infrared, by recognizing two main sources of non-perturbative effects: the strong coupling and the collinear PDFs. Both are evaluated at the low scale $q_T$ within the properly resummed TMD factorized cross sections, and are therefore probed at energies well below 1–2 GeV.
Of course, one could adopt completely different functional forms in this region, independent of the perturbative expressions, as long as they match them at larger $q_T$. The perspective adopted here is different: while perturbative QCD cannot be fully trusted at such low scales, the analytic structure it suggests is still considered meaningful. Accordingly, the behavior of $\alpha_S(q_T)$ in the deep infrared is modeled by acting on the logarithms $\log(Q/q_T)$ as in Eq.~\eqref{eq:alphaS_model}, which have been traded for the coupling via the renormalization group. Similarly, the PDFs are modeled by defining the DGLAP splitting kernels in the deep infrared, effectively parametrizing a backward non-perturbative evolution.
The parametrization of these non-perturbative models is ultimately to be guided by phenomenology on low-energy data. However, a full phenomenological implementation is beyond the scope of this work. The purpose of this paper is not to perform a detailed data analysis, but rather to introduce a new strategy for treating transverse-momentum-dependent distributions consistently across perturbative and non-perturbative regimes. 
Nevertheless, a simple parametrization such as Eq.\eqref{eq:alphaS_model}, together with a crude toy model for the PDFs where the DGLAP evolution is simply frozen below $Q_0$, already shows the potential of this perspective. First, the parameters introduced are sensible features of QCD, such as the effective gluon mass used to freeze the coupling at low scales. Secondly, the flavor dependence in the TMDs emerges naturally through the parametrization of the DGLAP evolution below 1–2 GeV, as shown in Fig.~\ref{fig:collpdfs}.
Once these parametrizations have been tested and validated at the level of cross sections, they can be consistently re-implemented in the $b_T$-space expressions. This allows for a clean extraction of the TMDs within the traditional formalism, fully leveraging perturbative QCD while keeping all non-perturbative effects under control, thereby reconciling the present $q_T$-space perspective with the historical approach.

Another key consequence of the framework proposed here is the central role of collinear PDFs in transverse momentum physics. Together with the strong coupling, they constitute the cornerstone around which cross sections and TMDs are constructed. This opens the way for a consistent extraction of PDFs from transverse momentum–dependent data, a task that remains challenging in standard approaches due to the proliferation of non-perturbative inputs and their entanglement with resummation prescriptions. Within the present perspective, the PDFs would constitute the only non-perturbative content for $q_T$ above $Q_0$. Input functions can be defined at the scale $Q_0$ and subsequently evolved perturbatively to high energies and non-perturbatively backwards.
Exploiting observables differential in $q_T$ to determine PDFs provides the invaluable advantage of probing regions typically inaccessible to collinear factorization, such as extreme $x$ values and very low energy scales. For instance, it has been suggested\cite{Gonzalez-Hernandez:2018ipj} that tensions in describing the large-$q_T$ tails in low-energy SIDIS may stem from poorly constrained PDFs in the relevant kinematics. A particularly relevant case is the gluon PDF, which is the most elusive and least constrained parton distribution: it contributes already at leading order at large $q_T$ and dominates gluon-initiated processes such as Higgs production.
In this regard, the present framework has been primarily devised for quark TMDs, but its extension to gluons is straightforward.

In conclusion, the strategy and formalism presented here represent a paradigm shift in TMD phenomenology, where the goal goes beyond achieving an excellent description of experimental data, and is to probe the very foundations of strong interactions, using transverse-momentum physics as a sharp tool to explore not just the structure of hadrons, but the deep-infrared regime of QCD.

\begin{acknowledgements}
I warmly thank Matteo Cerutti, Osvaldo Gonzalez-Hernandez, and Tommaso Rainaldi for stimulating discussions and helpful comments. I owe special thanks to Ted Rogers, whose critical insights were crucial for this work. I am also grateful to Alessandro Bacchetta  and Mariaelena Boglione for their comments and valuable feedback. I am further indebted to Marco Bonvini and Paolo Torrielli for insightful advice on analytic resummation methods.
This work was supported by the Italian Ministry of University and Research (MUR) grant PRIN 2022SNA23K funded by the European Union -- Next Generation EU, Mission 4, Component 2, CUP I53D23001410006.
\end{acknowledgements}

\appendix

\section{Functions required for Resummation}
\label{app:resummation}

In this Appendix, we collect the fixed order results and the functions participating in resummation up to NLL. 
The Sudakov factor up to this accuracy is completely specified by the functions
\begin{subequations}
\label{eq:g_functions}
    \begin{align}
    &g_1(\lambda) = \frac{\gamma_K^{[0]}}{2\beta_0} \Big(
    1 + \frac{\log{(1-\lambda)}}{\lambda}
    \Big)
    \label{eq:g1}
    \\
    &g_2(\lambda) = \frac{\gamma_K^{[0]}}{4 \beta_0^2} 
    \frac{\beta_1}{\beta_0} \frac{\lambda}{1-\lambda}
    \Big(1 + \frac{\log{(1-\lambda)}}{\lambda} + 
    \frac{1}{2}\frac{1-\lambda}{\lambda} \log^2{(1-\lambda)}\Big)
    \notag \\
    &\quad
    -\frac{\gamma_K^{[1]}}{4 \beta_0^2} 
    \Big( \frac{\lambda}{1-\lambda}  + \log{(1-\lambda)}\Big)
    -\frac{\gamma_f^{[0]}}{2\beta_0} \log{(1-\lambda)}
    \label{eq:g2}
    \end{align}
\end{subequations}
and the Collins-Soper kernel contributes:
\begin{subequations}
\label{eq:k_functions}
    \begin{align}
    &\kappa_1(\lambda) = \frac{\gamma_K^{[0]}}{2\beta_0} \log{(1-\lambda)}
    \\
    &\kappa_2(\lambda) = \frac{\gamma_K^{[1]}}{4 \beta_0^2} 
    \frac{\beta_1}{\beta_0} \frac{\lambda}{1-\lambda}
    \Big(\lambda + \log{(1-\lambda)}\Big)
    -\frac{\gamma_K^{[1]}}{4\beta_0^2} \frac{\lambda^2}{1-\lambda}
    \end{align}
\end{subequations}
The anomalous dimensions and the QCD Beta function coefficients are known and given by:
\begin{subequations}
\label{eq:anomalous_dim}
\begin{align}
    &\gamma_K^{[0]}  = 8 C_F, 
    & & \gamma_K^{[1]}  = \left(\frac{536}{9} - \frac{8 \pi^2}{3}\right) C_A C_F - \frac{80}{9} C_F N_f, \\
    &\gamma_f^{[0]}  = 6 C_F, & & \\ 
    &\beta_0  = \frac{11}{3} C_A - \frac{2}{3} N_f, 
    & &\beta_1  = \frac{34}{3} C_A^2 - \frac{10}{3} C_A N_f - 2 C_F N_f.
\end{align}
\end{subequations}
where $C_F = 4/3$ and $C_A = 3$, while $N_f$ varies with energy, increasing by one unit at each heavy quark mass threshold crossing. The convention here is that perturbative expansions are defined in terms of $\alpha_S/(4\pi)$.
Finally, we provide the DGLAP splitting kernels at lowest order entering into the OPE of quark TMDs at large transverse momentum:
\begin{subequations}
    \label{eq:splitting_kernel_LO}
    \begin{align}
    &P^{[0]}_{q_j/q_k}(x) = \delta_{j k} 2 C_F \Big[ \frac{1+x^2}{(1-x)_+} + \frac{3}{2}\delta(1-x) \Big],\\
    &P^{[0]}_{q_j/g}(x) = x^2 + (1-x)^2.
    \end{align}
\end{subequations}
with $P^{[0]}_{\overline{q}_j/\overline{q}_k} = P^{[0]}_{q_j/q_k}$ and $P^{[0]}_{q_j/\overline{q}_k} = 0$; and the non-logarithmic part of the Wilson coefficient for the OPE at small-$b_T$:
\begin{subequations}
    \label{eq:Wilson_coeffs_NLO}
    \begin{align}
    &C^{[1]}_{q_j/q_k}(x) = \delta_{j k}\,2 C_F \Big[1-x - \frac{\pi^2}{12}\delta(1-x)  \Big],\\
    &C^{[1]}_{q_j/g}(x) = 2 x\,(1-x)
    \end{align}
\end{subequations}
with $C^{[1]}_{\overline{q}_j/\overline{q}_k} = C^{[1]}_{q_j/q_k}$ and $C^{[1]}_{q_j/\overline{q}_k} = 0$;

\section{Analytic properties of functions at NLL}
\label{app:analytic_NLL}

In this Section, we discuss some useful analytic properties of the functions contributing to the resummation in transverse momentum space at NLL. 
First, consider the boundary term in Eq.\eqref{eq:I_above}, i.e., the contribution associated with the hypergeometric function. It depends solely on $h_1$ and will henceforth be denoted by $B_{c_1}$. 
It can be re-written as:
\begin{align}
    \label{eq:boundary_prop1}
    &B_{c_1}(h_1) = 
    J_0(c_1) - {}_1F_2\left(\frac{h_1}{2};1,1+\frac{h_1}{2};-\frac{c_1^2}{4}\right).
\end{align}
Moreover, its behavior in the limits of small and large $h_1$ can be expressed as:
\begin{align}
\label{eq:boundary_prop2}
&B_{c_1}(h_1) =
\notag \\
&\quad
\begin{cases}
J_0(c_1) - 1 - \sum_{n\geq0}
\left(-\frac{h_1}{2}\right)^{n+1} \frac{c_1^2}{4}
{}_{n+2}F_{n+3}\Bigl(1,\dots;2,\dots;-\frac{c_1^2}{4}\Bigr), & |h_1| < 2 \\[1em]
- \frac{c_1}{h_1} J_1(c_1) + \sum_{n\geq0}
\left(-\frac{2}{h_1}\right)^{n+2} \frac{c_1^2}{4}
{}_{n}F_{n+1}\Bigl(2,\dots;1,\dots;-\frac{c_1^2}{4}\Bigr), & |h_1| > 2
\end{cases}
\end{align}
where the $\dots$ indicate repeated occurrences of the same value.
Note that the transition point $|h_1| = 2$ corresponds to the first pole of Eq.\eqref{eq:Psi0}. 
Finally, its derivatives with respect to $h_1$ are also expressible in terms of hypergeometric functions:
\begin{align}
    \label{eq:boundary_prop3}
    &\frac{d^n\,B_{c_1}(h_1)}{d h_1^n} = 
    \frac{c_1^2}{4}\frac{n!}{2^n}\left(-\frac{1}{1+\frac{h_1}{2}}\right)^{n+1}
        {}_{n+1}F_{n+2}\left(1+\frac{h_1}{2},\dots;2,2+\frac{h_1}{2},\dots;-\frac{c_1^2}{4}\right).
\end{align}

\bigskip

Next, we focus on the properties of the series in Eq.\eqref{eq:I_below}. Its asymptotic behavior at large $A \sim 1/a_S(\mu)$ depends on the sign of $(z_n - 1)$, where $z_n = 1$ marks the same poles of the function $\Psi_0$ in Eq.\eqref{eq:Psi0}.
In the OPE region, $h_1$ is small and positive, and thus $z_n > 1$. 
In this regime, the following asymptotic expansion holds:
\begin{align}
\label{eq:zn>1_asy}
    &(A\,z_n)^{-A} e^{A z_n} \Gamma_{1+A}(A \, z_n) 
    = 
    \sum_{k\geq 0}  \frac{1}{z_n^k} 
    \sum_{j\geq 0} \frac{(-k)_j}{j!} B_j^{(k+1)}(k) \frac{1}{A^j}
\end{align}
where $B_n^{(a)}(x)$ is the generalized Bernoulli polynomial of degree $n$. 
Upon inserting this expansion, the function $\Phi$ in Eq.\eqref{eq:Phi} reads:
\begin{align}
    \label{eq:Phi_asyOPE}
    &\Phi(\lambda_c, L_c) = 
    -\frac{1}{2 L} \frac{h_1(\lambda)}{1-\lambda}
    \frac{d^2}{d h_1^2} B_{c_1}(h_1)
    +
    \ordof{1/L^2}\,.
\end{align}
The $j=0$ term in Eq.\eqref{eq:zn>1_asy} cancels exactly the boundary contribution of Eq.\eqref{eq:I_above}. The leading contribution then comes from the $j=1$ term, which is of NNLL order, i.e. beyond the accuracy of our calculation. For $k_c \approx \mu$, this evaluates to:
\begin{align}
    \label{eq:Phi_asyOPE_aS}
    &\Phi = 
    - a_S(\mu) C_F \times (0.618\dots)
    +
    \mathcal{O}(a_S(\mu)^2 L_c)
\end{align}
confirming that it contributes only at NNLL order.

In the opposite limit of low transverse momentum, $h_1$ becomes large and negative, and thus to $z_n < 1$. The asymptotic behavior in this region is determined by
\begin{align}
    \label{eq:zn<1}
    &(A\,z_n)^{-A} e^{A z_n} \Gamma_{1+A}(A \, z_n) 
    =
    e^{-A(1-z_n + \log{z_n})}
    \sqrt{2 \pi A} \left[ 
    1+\mathcal{O}\left(\frac{1}{A}\right)\right] \,.
\end{align}
Because of this behavior, the series in Eq.\eqref{eq:I_below} develops an enhancement that perfectly cancels out the leading Sudakov suppression $e^{L_c g_1}$:

\begin{align}
    &\text{series in Eq.\eqref{eq:I_below}}
   = e^{-L_c g_1(\lambda_c)}
   \sqrt{8 \pi c \frac{L_c}{\lambda_c}}\,
   \frac{c_1 e^{-L_c}}{4} J_1 \Big(\frac{c_1 e^{-L_c}}{2}\Big)
    + \mathcal{O}(1/L_c)
    \notag \\
    &\quad
    =e^{-L_c g_1(\lambda_c)}
   \frac{c_1^2}{16} 
   \sqrt{\frac{\gamma_K^{[0]}}{4\beta_0^2} \frac{8\pi}{a_S(\mu)}}\,
   \frac{k_c^2}{\mu^2}
    + \mathcal{O}(1/L_c,k_c^4/\mu^2).
\end{align}
This result is particularly significant: it explicitly shows the breakdown of the LL Sudakov suppression, while simultaneously reproducing the 
quadratic behavior first identified by Parisi and Petronzio~\cite{Parisi:1979se}, which is missed at LL accuracy. 

\section{Dependence on the rapidity scale in transverse momentum space}
\label{app:zeta_rap}

The rapidity scale $\zeta$ was not explicitly addressed in Section\ref{sec:analytic_res}, as it was most often set to $\zeta = \mu^2$. This choice removes the genuine Collins–Soper kernel contribution in the last line of Eq.\eqref{eq:tmd_bT_resummed}. The motivation behind this simplification is that, in most applications, the dependence on $\zeta$ eventually cancels out at the level of cross sections, and can therefore be safely neglected in intermediate steps. 
In this appendix we show how to retain the full $\zeta$-dependence in the analytic Fourier inversion, thereby generalizing the NLL result of Eq.\eqref{eq:F_NLL}. 

In the integrand of the cumulative distribution $R_j$ at NLL, Eq.~\eqref{eq:R_NLL_prelim}, the inclusion of the rapidity scale dependence requires only two simple modifications:
\begin{itemize}
    \item The NLL Collins–Soper kernel is added to the Sudakov factor,
    \begin{align}
        S^{(NLL)}(\lambda_c, L_c)
        \to S^{(NLL)}(\lambda_c, L_c)
        +
       \frac{1}{2} L_\zeta K^{(NLL)}(\lambda_c,L_c)
    \end{align}
    with
    \begin{align}
        \label{eq:CS_kernel_NLL}
        &K^{(NLL)}(\lambda_c,L_c) = \kappa_1(\lambda_c) + \frac{1}{L_c}\kappa_2(\lambda_c)
    \end{align}
    \item The function $H_1$ acquires an additional contribution induced by the Collins–Soper kernel,
    \begin{align}
        &H_1(\lambda_c, {L_\xi}/{L_c}\big) \to 
        H_1(\lambda_c, {L_\xi}/{L_c}\big) + \frac{1}{2}L_\zeta \mathcal{H}_1\big(\lambda_c, {L_\xi}/{L_c}\big)
    \end{align}
    with $\mathcal{H}_1$ coinciding with $H_2$ of Eq.~\eqref{eq:H2} upon replacing $h_2$ by $\kappa_1$:
\begin{align}
    \label{eq:H1CS_explicit}
    &\mathcal{H}_1(\lambda_c,r) = \frac{\gamma_K^{[0]}}{2\beta_0} \log{\Big(1 - \frac{\lambda_c}{1-\lambda_c} r\Big)}.
\end{align}
\end{itemize}
These replacements result into
\begin{align}
    \label{eq:R_NLL_prelim_zeta}
    &R^{(NLL)}_j\big(x,k_c,\mu,\zeta/\mu^2\big)
    =
    e^{L_c g_1(\lambda_c) + g_2(\lambda_c) + \frac{1}{2} L_\zeta 
    K^{(NLL)}(\lambda_c,L_c)}
    f_j(x;k_c)
    \notag \\
    &\times
    \int_0^\infty d\xi J_1(\xi) 
    e^{L_\xi H_1\big(\lambda_c, {L_\xi}/{L_c}\big)
    +
    \frac{1}{2}L_\zeta
    \mathcal{H}_1\big(\lambda_c, {L_\xi}/{L_c}\big)}.
\end{align}
The integration restricted to $\xi \geq c_1$ is solved in saddle–point approximation, leading to the same result as Eq.~\eqref{eq:I_above} with $h_1$ replaced by $h_1 \big(1 + L_\zeta/(2L_c)\big)$.
On the other hand, the integration for $\xi \leq c_1$ yields the same result as Eq.~\eqref{eq:I_below}, with the sole substitution $A \to A + c L_\zeta/2$. 

With these substitutions, the final result easily follows from the analogous equations obtained in Section\ref{sec:analytic_res}. It is important to stress that in the cumulative distribution $R_j$, when evaluated at $k_c = \mu$, the entire dependence on the rapidity scale is sub-leading compared to NLL. 
This is important, as $R_j$ in this form provides an alternative renormalization scheme\cite{Gonzalez-Hernandez:2022ifv} for collinear PDFs, which are independent of $\zeta$. It is less straightforward, but merely a matter of calculation, to show that the TMD distributions depend on $L_\zeta$ according to the Collins-Soper evolution after the analytic Fourier transform.

\bibliographystyle{unsrt}
\bibliography{bibliography}

\begin{thebibliography}{10}

\bibitem{Collins:2011zzd}
John Collins.
\newblock {\em {Foundations of Perturbative QCD}}, volume~32.
\newblock Cambridge University Press, 2011.

\bibitem{Simonelli:2025xpm}
Andrea Simonelli, Alberto Accardi, Matteo Cerutti, Caroline S.~R. Costa, and
  Andrea Signori.
\newblock {Unveiling the Collins-Soper kernel in inclusive DIS at threshold}.
\newblock 2 2025.

\bibitem{Parisi:1979se}
G.~Parisi and R.~Petronzio.
\newblock {Small Transverse Momentum Distributions in Hard Processes}.
\newblock {\em Nucl. Phys. B}, 154:427--440, 1979.

\bibitem{Collins:1981uk}
John~C. Collins and Davison~E. Soper.
\newblock {Back-To-Back Jets in QCD}.
\newblock {\em Nucl. Phys. B}, 193:381, 1981.
\newblock [Erratum: Nucl.Phys.B 213, 545 (1983)].

\bibitem{Collins:1981va}
John~C. Collins and Davison~E. Soper.
\newblock {Back-To-Back Jets: Fourier Transform from B to K-Transverse}.
\newblock {\em Nucl. Phys. B}, 197:446--476, 1982.

\bibitem{Collins:1984kg}
John~C. Collins, Davison~E. Soper, and George~F. Sterman.
\newblock {Transverse Momentum Distribution in Drell-Yan Pair and W and Z Boson
  Production}.
\newblock {\em Nucl. Phys. B}, 250:199--224, 1985.

\bibitem{Ellis:1997ii}
R.~Keith Ellis and Sinisa Veseli.
\newblock {$W$ and $Z$ transverse momentum distributions: Resummation in
  $q_{T}$ space}.
\newblock {\em Nucl. Phys. B}, 511:649--669, 1998.

\bibitem{Ellis:1997sc}
R.~Keith Ellis, D.~A. Ross, and Sinisa Veseli.
\newblock {Vector boson production in hadronic collisions}.
\newblock {\em Nucl. Phys. B}, 503:309--338, 1997.

\bibitem{Frixione:1998dw}
Stefano Frixione, Paolo Nason, and Giovanni Ridolfi.
\newblock {Problems in the resummation of soft gluon effects in the transverse
  momentum distributions of massive vector bosons in hadronic collisions}.
\newblock {\em Nucl. Phys. B}, 542:311--328, 1999.

\bibitem{Kulesza:1999gm}
Anna Kulesza and W.~James Stirling.
\newblock {Sudakov logarithm resummation in transverse momentum space for
  electroweak boson production at hadron colliders}.
\newblock {\em Nucl. Phys. B}, 555:279--305, 1999.

\bibitem{Kulesza:1999sg}
Anna Kulesza and W.~James Stirling.
\newblock {On the resummation of subleading logarithms in the transverse
  momentum distribution of vector bosons produced at hadron colliders}.
\newblock {\em JHEP}, 01:016, 2000.

\bibitem{Kulesza:2001jc}
Anna Kulesza and W.~James Stirling.
\newblock {Soft gluon resummation in transverse momentum space for electroweak
  boson production at hadron colliders}.
\newblock {\em Eur. Phys. J. C}, 20:349--356, 2001.

\bibitem{Ebert:2016gcn}
Markus~A. Ebert and Frank~J. Tackmann.
\newblock {Resummation of Transverse Momentum Distributions in Distribution
  Space}.
\newblock {\em JHEP}, 02:110, 2017.

\bibitem{Kang:2017cjk}
Daekyoung Kang, Christopher Lee, and Varun Vaidya.
\newblock {A fast and accurate method for perturbative resummation of
  transverse momentum-dependent observables}.
\newblock {\em JHEP}, 04:149, 2018.

\bibitem{Banfi:2004yd}
Andrea Banfi, Gavin~P. Salam, and Giulia Zanderighi.
\newblock {Principles of general final-state resummation and automated
  implementation}.
\newblock {\em JHEP}, 03:073, 2005.

\bibitem{Banfi:2014sua}
Andrea Banfi, Heather McAslan, Pier~Francesco Monni, and Giulia Zanderighi.
\newblock {A general method for the resummation of event-shape distributions in
  $e^{+} e^{-}$ annihilation}.
\newblock {\em JHEP}, 05:102, 2015.

\bibitem{Monni:2016ktx}
Pier~Francesco Monni, Emanuele Re, and Paolo Torrielli.
\newblock {Higgs Transverse-Momentum Resummation in Direct Space}.
\newblock {\em Phys. Rev. Lett.}, 116(24):242001, 2016.

\bibitem{Gonzalez-Hernandez:2022ifv}
J.~O. Gonzalez-Hernandez, T.~C. Rogers, and N.~Sato.
\newblock {Combining nonperturbative transverse momentum dependence with TMD
  evolution}.
\newblock {\em Phys. Rev. D}, 106(3):034002, 2022.

\bibitem{Gonzalez-Hernandez:2023iso}
J.~O. Gonzalez-Hernandez, T.~Rainaldi, and T.~C. Rogers.
\newblock {Resolution to the problem of consistent large transverse momentum in
  TMDs}.
\newblock {\em Phys. Rev. D}, 107(9):094029, 2023.

\bibitem{Moch:2004pa}
S.~Moch, J.~A.~M. Vermaseren, and A.~Vogt.
\newblock {The Three loop splitting functions in QCD: The Nonsinglet case}.
\newblock {\em Nucl. Phys. B}, 688:101--134, 2004.

\bibitem{Vogt:2004mw}
A.~Vogt, S.~Moch, and J.~A.~M. Vermaseren.
\newblock {The Three-loop splitting functions in QCD: The Singlet case}.
\newblock {\em Nucl. Phys. B}, 691:129--181, 2004.

\bibitem{Collins:2017oxh}
John Collins and Ted~C. Rogers.
\newblock {Connecting Different TMD Factorization Formalisms in QCD}.
\newblock {\em Phys. Rev. D}, 96(5):054011, 2017.

\bibitem{Henn:2019swt}
Johannes~M. Henn, Gregory~P. Korchemsky, and Bernhard Mistlberger.
\newblock {The full four-loop cusp anomalous dimension in $\mathcal{N}=4$ super
  Yang-Mills and QCD}.
\newblock {\em JHEP}, 04:018, 2020.

\bibitem{Bailey:2020ooq}
S.~Bailey, T.~Cridge, L.~A. Harland-Lang, A.~D. Martin, and R.~S. Thorne.
\newblock {Parton distributions from LHC, HERA, Tevatron and fixed target data:
  MSHT20 PDFs}.
\newblock {\em Eur. Phys. J. C}, 81(4):341, 2021.

\bibitem{Catani:1996yz}
Stefano Catani, Michelangelo~L. Mangano, Paolo Nason, and Luca Trentadue.
\newblock {The Resummation of soft gluons in hadronic collisions}.
\newblock {\em Nucl. Phys. B}, 478:273--310, 1996.

\bibitem{Kulesza:2002rh}
Anna Kulesza, George~F. Sterman, and Werner Vogelsang.
\newblock {Joint resummation in electroweak boson production}.
\newblock {\em Phys. Rev. D}, 66:014011, 2002.

\bibitem{Bonvini:2008ei}
Marco Bonvini, Stefano Forte, and Giovanni Ridolfi.
\newblock {Borel resummation of transverse momentum distributions}.
\newblock {\em Nucl. Phys. B}, 808:347--363, 2009.

\bibitem{Bacchetta:2022awv}
Alessandro Bacchetta, Valerio Bertone, Chiara Bissolotti, Giuseppe Bozzi,
  Matteo Cerutti, Fulvio Piacenza, Marco Radici, and Andrea Signori.
\newblock {Unpolarized transverse momentum distributions from a global fit of
  Drell-Yan and semi-inclusive deep-inelastic scattering data}.
\newblock {\em JHEP}, 10:127, 2022.

\bibitem{Moos:2023yfa}
Valentin Moos, Ignazio Scimemi, Alexey Vladimirov, and Pia Zurita.
\newblock {Extraction of unpolarized transverse momentum distributions from the
  fit of Drell-Yan data at N$^{4}$LL}.
\newblock {\em JHEP}, 05:036, 2024.

\bibitem{Bacchetta:2024qre}
Alessandro Bacchetta, Valerio Bertone, Chiara Bissolotti, Giuseppe Bozzi,
  Matteo Cerutti, Filippo Delcarro, Marco Radici, Lorenzo Rossi, and Andrea
  Signori.
\newblock {Flavor dependence of unpolarized quark transverse momentum
  distributions from a global fit}.
\newblock {\em JHEP}, 08:232, 2024.

\bibitem{Moos:2025sal}
Valentin Moos, Ignazio Scimemi, Alexey Vladimirov, and Pia Zurita.
\newblock {Determination of unpolarized TMD distributions from the fit of
  Drell-Yan and SIDIS data at N$^4$LL}.
\newblock 3 2025.

\bibitem{Bacchetta:2025ara}
Alessandro Bacchetta, Valerio Bertone, Chiara Bissolotti, Matteo Cerutti, Marco
  Radici, Simone Rodini, and Lorenzo Rossi.
\newblock {Neural-Network Extraction of Unpolarized
  Transverse-Momentum-Dependent Distributions}.
\newblock {\em Phys. Rev. Lett.}, 135(2):021904, 2025.

\bibitem{Camarda:2025lbt}
Stefano Camarda, Giancarlo Ferrera, and Lorenzo Rossi.
\newblock {Drell--Yan lepton pair production at low invariant masses:
  transverse-momentum resummation and non-perturbative effects in QCD}.
\newblock 8 2025.

\bibitem{Cerutti:2025inprep}
M.~Cerutti and A~Simonelli.
\newblock {\em In preparation}.

\bibitem{Bizon:2017rah}
Wojciech Bizon, Pier~Francesco Monni, Emanuele Re, Luca Rottoli, and Paolo
  Torrielli.
\newblock {Momentum-space resummation for transverse observables and the Higgs
  p$_{\perp}$ at N$^{3}$LL+NNLO}.
\newblock {\em JHEP}, 02:108, 2018.

\bibitem{Re:2021con}
Emanuele Re, Luca Rottoli, and Paolo Torrielli.
\newblock {Fiducial Higgs and Drell-Yan distributions at N$^3$LL$^\prime$+NNLO
  with RadISH}.
\newblock {\em JHEP}, 2109:108, 2021.

\bibitem{Buonocore:2024xmy}
Luca Buonocore, Luca Rottoli, and Paolo Torrielli.
\newblock {Resummation of combined QCD-electroweak effects in Drell Yan
  lepton-pair production}.
\newblock {\em JHEP}, 07:193, 2024.

\bibitem{Scimemi:2017etj}
Ignazio Scimemi and Alexey Vladimirov.
\newblock {Analysis of vector boson production within TMD factorization}.
\newblock {\em Eur. Phys. J. C}, 78(2):89, 2018.

\bibitem{Scimemi:2018xaf}
Ignazio Scimemi and Alexey Vladimirov.
\newblock {Systematic analysis of double-scale evolution}.
\newblock {\em JHEP}, 08:003, 2018.

\bibitem{Catani:1992ua}
S.~Catani, L.~Trentadue, G.~Turnock, and B.~R. Webber.
\newblock {Resummation of large logarithms in e+ e- event shape distributions}.
\newblock {\em Nucl. Phys. B}, 407:3--42, 1993.

\bibitem{delRio:2024vvq}
Oscar del Rio, Alexei Prokudin, Ignazio Scimemi, and Alexey Vladimirov.
\newblock {Transverse momentum moments}.
\newblock {\em Phys. Rev. D}, 110(1):016003, 2024.

\bibitem{Simonelli:2024vyh}
Andrea Simonelli.
\newblock {Analytic solutions of the DGLAP evolution and theoretical
  uncertainties}.
\newblock {\em Eur. Phys. J. C}, 84(8):867, 2024.

\bibitem{Aglietti:2025ezs}
Ugo~Giuseppe Aglietti, Giancarlo Ferrera, and Wan-Li Ju.
\newblock {Saddle-point method for resummed form factors in QCD}.
\newblock 6 2025.

\bibitem{Aslan:2024nqg}
F.~Aslan, M.~Boglione, J.~O. Gonzalez-Hernandez, T.~Rainaldi, T.~C. Rogers, and
  A.~Simonelli.
\newblock {Phenomenology of TMD parton distributions in Drell-Yan and Z0 boson
  production in a hadron structure oriented approach}.
\newblock {\em Phys. Rev. D}, 110(7):074016, 2024.

\bibitem{Ito:1980ev}
A.~S. Ito et~al.
\newblock {Measurement of the Continuum of Dimuons Produced in High-Energy
  Proton - Nucleus Collisions}.
\newblock {\em Phys. Rev. D}, 23:604--633, 1981.

\bibitem{Harland-Lang:2014zoa}
L.~A. Harland-Lang, A.~D. Martin, P.~Motylinski, and R.~S. Thorne.
\newblock {Parton distributions in the LHC era: MMHT 2014 PDFs}.
\newblock {\em Eur. Phys. J. C}, 75(5):204, 2015.

\bibitem{Buckley:2014ana}
Andy Buckley, James Ferrando, Stephen Lloyd, Karl Nordstr{\"o}m, Ben Page,
  Martin R{\"u}fenacht, Marek Sch{\"o}nherr, and Graeme Watt.
\newblock {LHAPDF6: parton density access in the LHC precision era}.
\newblock {\em Eur. Phys. J. C}, 75:132, 2015.

\bibitem{CDF:1999bpw}
T.~Affolder et~al.
\newblock {The transverse momentum and total cross section of $e^+e^-$ pairs in
  the $Z$ boson region from $p\bar{p}$ collisions at $\sqrt{s} = 1.8$ TeV}.
\newblock {\em Phys. Rev. Lett.}, 84:845--850, 2000.

\bibitem{Rogers:2024cci}
Ted Rogers, F.~Aslan, M.~Boglione, J.~O. Gonzalez-Hernandez, T.~Rainaldi, and
  A.~Simonelli.
\newblock {TMD phenomenology motivated by nonperturbative structures}.
\newblock {\em PoS}, Transversity2024:030, 2024.

\bibitem{Mattingly:1993ej}
A.~C. Mattingly and Paul~M. Stevenson.
\newblock {Optimization of R(e+ e-) and 'freezing' of the QCD couplant at
  low-energies}.
\newblock {\em Phys. Rev. D}, 49:437--450, 1994.

\bibitem{Deur:2016tte}
Alexandre Deur, Stanley~J. Brodsky, and Guy~F. de~Teramond.
\newblock {The QCD Running Coupling}.
\newblock {\em Nucl. Phys.}, 90:1, 2016.

\bibitem{Deur:2005cf}
A.~Deur, V.~Burkert, Jian-Ping Chen, and W.~Korsch.
\newblock {Experimental determination of the effective strong coupling
  constant}.
\newblock {\em Phys. Lett. B}, 650:244--248, 2007.

\bibitem{Deur:2008rf}
A.~Deur, V.~Burkert, J.~P. Chen, and W.~Korsch.
\newblock {Determination of the effective strong coupling constant
  alpha(s,g(1))(Q**2) from CLAS spin structure function data}.
\newblock {\em Phys. Lett. B}, 665:349--351, 2008.

\bibitem{Grunberg:1980ja}
G.~Grunberg.
\newblock {Renormalization Group Improved Perturbative QCD}.
\newblock {\em Phys. Lett. B}, 95:70, 1980.
\newblock [Erratum: Phys.Lett.B 110, 501 (1982)].

\bibitem{Brodsky:1994eh}
Stanley~J. Brodsky and Hung~Jung Lu.
\newblock {Commensurate scale relations in quantum chromodynamics}.
\newblock {\em Phys. Rev. D}, 51:3652--3668, 1995.

\bibitem{Deur:2004ti}
A.~Deur et~al.
\newblock {Experimental determination of the evolution of the Bjorken integral
  at low $Q^2$}.
\newblock {\em Phys. Rev. Lett.}, 93:212001, 2004.

\bibitem{Kim:1998kia}
J.~H. Kim et~al.
\newblock {A Measurement of alpha(s)(Q**2) from the Gross-Llewellyn Smith sum
  rule}.
\newblock {\em Phys. Rev. Lett.}, 81:3595--3598, 1998.

\bibitem{Gonzalez-Hernandez:2018ipj}
J.~O. Gonzalez-Hernandez, T.~C. Rogers, N.~Sato, and B.~Wang.
\newblock {Challenges with Large Transverse Momentum in Semi-Inclusive Deeply
  Inelastic Scattering}.
\newblock {\em Phys. Rev. D}, 98(11):114005, 2018.

\end{thebibliography}

\end{document}